\documentclass[12pt]{article}
\usepackage{amssymb}
\begin{document}

\begin{flushright}
{\tt hep-th/0411164}
\end{flushright}

\vspace{5mm}

\begin{center}
{{{\Large \bf Domain Walls in Noncommutative Field Theories}}\\[18mm]
{Yoonbai Kim,~~O-Kab Kwon,~~Chong Oh Lee}\\[3mm]
{\it BK21 Physics Research Division and Institute of
Basic Science,\\
Sungkyunkwan University, Suwon 440-746, Korea}\\
{\tt yoonbai@skku.edu~~okab@skku.edu~~cohlee@newton.skku.ac.kr} }
\end{center}
\vspace{10mm}

\begin{abstract}
We study kinks in a wide class of noncommutative (NC) field
theories. We find rich structure of the static kinks in DBI type
NC tachyon action for an unstable D$p$-brane with
general constant open string metric and NC parameter.
Among which thick topological BPS NC kink and tensionless
half-kink are particularly intriguing. Reproduction of the correct
decent relation between D$p$ and D$(p-1)$ lets us interpret the
obtained NC kinks as codimension-one D-brane and its composites.
If we turn on DBI type NC U(1) gauge field on an
unstable D2-brane, only constant field strength is allowed by
gauge equation and NC Bianchi identity. Inclusion of the NC U(1)
gauge field induces fundamental string charge localized on the
codimension-one brane, which turns D$(p-1)$ into D$(p-1)$F1.
\end{abstract}

\newpage

\setcounter{equation}{0}
\section{Introduction}

Dynamics of an unstable D-brane or D-brane-anti-D-brane in string theory
has attracted much attention~\cite{Sen:2004nf}. Instability of such system is
represented by tachyonic degree and its dynamics is depicted by
condensation of the tachyon field and related phenomena. Recently
real-time description of the homogeneous tachyon condensation has been
made by finding rolling tachyon solutions in both
boundary conformal field theory (BCFT)
and effective field theory (EFT)~\cite{Sen:2002nu,Sen:2002an}.
When the tachyon is condensed,
the list of questions involves dynamical description of decaying
processes including inhomogeneity~\cite{Cline:2002it}
and the formation of final products after
the unstable D-brane decays.
For the latter question, either perturbative degrees namely
closed string degrees are
produced~\cite{Lambert:2003zr,Kluson:2003qk}
or nonperturbative lower-dimensional
branes are formed, in which the important objects are
stable codimension-one D-branes or their
composites including confined fundamental string
(F1)~\cite{Lambert:2003zr,Kim:2003in,Sen:2003bc}.

Tachyon dynamics has been studied by using both various string
theoretic methods including BCFT, open string field theory,
boundary string field theory (BSFT), $c=1$ matrix model, and
effective field theoretic languages including EFT, noncommutative
field theory (NCFT), $p$-adic string theory. When constant
Neveu-Schwarz (NS) antisymmetric two-form field exists, an
effective description of string theory is to employ the
NCFT~\cite{Seiberg:1999vs}. Naturally dynamics of the unstable
D-brane by using noncommutative (NC) tachyon condensation has also
been an attractive topic from the
beginning~\cite{Dasgupta:2000ft,Harvey:2000jt,Gopakumar:2000rw}.
On the production of lower-dimensional branes, studies are mostly
about codimension-two objects since Gopakumar-Minwalla-Strominger
(GMS) solitons~\cite{Gopakumar:2000zd} and their
analogues~\cite{Polychronakos:2000zm} are basic solitonic objects
for the analysis. These provide a good description of D$(p-2)$s
from D$p{\bar {\rm D}}p$ system, however another representative
example is production of stable D$(p-1)$-branes from an unstable
D$p$-brane with or without both NS two-form field and
Dirac-Born-Infeld (DBI) type electromagnetism. Though there have
been many studies on both NC rolling
tachyon~\cite{Mukhopadhyay:2002en} and codimension-one branes in
terms of
EFT~\cite{Lambert:2003zr,Sen:2003tm,Kim:2003in,Brax:2003rs,Kim:2003ma,Sen:2003zf}
and BCFT~\cite{Sen:2003bc}, an NCFT study was recently performed
for D0s from an unstable D1 without and with DBI electric
field~\cite{Banerjee:2004cw}.

In this paper we study domain walls and tachyon kinks in general
NCFT settings and a DBI type NCFT including NC tachyon field. In
section 2, we consider NCFT actions of a real NC scalar field with
usual kinetic and polynomial type potential terms, and find all
possible NC domain wall configurations where many of them are
given by exact solutions. Extensions to the action with
nonpolynomial scalar potential and an EFT from
$p$-adic string theory are also briefly discussed. In section 3,
$(p+1)$-dimensional DBI type NC tachyon action in the background
of general constant open string metric and NC parameter with
runaway tachyon potential is taken into account. All possible
regular static codimension-one solitons are obtained, which are
classified by array of NC kink-antikink, single topological NC
kink which are either BPS or non-BPS due to open string metric,
tensionless NC half-kink, NC bounce, and hybrid of two NC
half-kinks. Since the decent relation between tension of $p$-brane
and $(p-1)$-brane is correctly reproduced, the obtained kinks are
interpreted as D$(p-1)$-branes and their composites. In section 4,
we turn on DBI type NC U(1) gauge field on a D2-brane in the
background of open string metric and NC parameter. Though the
analysis is complicated, every electric and magnetic component of
the NC field strength tensor on the D2-brane is proved to be
constant. The obtained NC tachyon kink configurations have the
same functional shapes as those in section 3, but their
interpretation leads to D$(p-1)$F1 and their composites, where F1s
are localized in D$(p-1)$. We conclude in section 5 with brief
summary and discussion.

\setcounter{equation}{0}
\section{Domain Walls in NC Scalar Field Theory}
Let us begin this section with a well-known local field theory
action of a real scalar field $\phi$ in (1+1)-dimensions
\begin{equation}\label{sac}
S_{{\rm EFT}2}=\int dtdx\left[-\frac{1}{2}\partial_{\mu}\phi\partial^{\mu}\phi
-V(\phi)\right],
\end{equation}
where the scalar potential $V(\phi)$ has at least two stable vacua, say
$\{\phi_{i}\, |\, dV/d\phi|_{\phi=\phi_{i}}=0,\;
d^{2}V/d\phi^{2}|_{\phi=\phi_{i}}>0,\;
~i=1,2,...\, \}$, with vanishing cosmological constant
$V(\phi_{i})=0$. It is also well-known that this model (\ref{sac}) supports
static topological kinks (domain walls) as classical solutions.
Here we briefly review how to obtain them.

Momentum density $T_{01}$ vanishes
for static configurations $\phi=\phi(x)$, and then pressure component
$T^{11}$ is rewritten as
\begin{eqnarray}
T^{11}=\frac{1}{2}\left(\phi'-\sqrt{2V(\phi)}\right)
\left(\phi'+\sqrt{2V(\phi)}\right),
\end{eqnarray}
where $\phi'=d\phi/dx$. A definition of Bogomol'nyi bound is given
by vanishing pressure everywhere, which provides a first-order
equation so-called BPS equation
\begin{equation}\label{BPS}
\phi'=\pm\sqrt{2V(\phi)}\, .
\end{equation}
For any static configuration satisfying the BPS equation (\ref{BPS})
saturates BPS bound, i.e., energy has minimum value bounded by
topological charge
\begin{eqnarray}\label{hen}
H&=&\int_{-\infty}^{\infty}dx \, T_{00}\\
&=&\int_{-\infty}^{\infty}dx\,\left[\frac{1}{2}\phi'^{2}
+V(\phi)\right] \\
&=&\int_{-\infty}^{\infty}dx\,\left[\frac{1}{2}\left(
\phi'\mp \sqrt{2V(\phi)}\right)^{2}\pm
Q(\phi)'\right] \\
&\ge &\pm \left[Q(\phi(\infty))-Q(\phi(-\infty))\right],
\label{toc}
\end{eqnarray}
where $Q'=\sqrt{2V}\phi'$ and $\phi(\pm\infty)$ is one of
$\phi_{i}$s. The equality in the last line (\ref{toc}) holds for
any BPS object, and the difference corresponds to topological
charge of a kink or an antikink in order of the signature in front
of the formula (\ref{toc}). Because of this nature, any static
solution satisfying first-order BPS equation (\ref{BPS}) also
satisfies second-order Euler-Lagrange equation automatically,
which can be derived through small variation of the energy
$H$~(\ref{hen}).

Simplest model is given by $\phi^{4}$-potential
\begin{equation}\label{spo}
V(\phi)=\frac{\lambda}{4}(\phi^{2}-v^{2})^{2}
\end{equation}
with degenerate superselective vacua of $i=1,2$ at $\phi_{i}=\pm
v$. Substituting the scalar potential (\ref{spo}) into the BPS
equation (\ref{BPS}) with boundary conditions, we obtain exact
kink (antikink) solution
\begin{equation}
\frac{\phi(x)}{v}=\pm \tanh \left[\frac{m_{{\rm H}}}{\sqrt{2}}(x-x_{0})\right],
\end{equation}
where $m_{{\rm H}}=\sqrt{\lambda}v$ is Higgs mass and an integration
constant $x_{0}$ is a zero mode.
The energy density $T_{00}$ of a kink (antikink) is localized near
its location $x_{0}$ with thickness $1/m_{{\rm H}}$
\begin{equation}
T_{00}=\frac{\lambda}{2}v^{4}{\rm sech}^{4}
\left[\frac{m_{{\rm H}}}{\sqrt{2}}(x-x_{0})\right],
\end{equation}
and its energy given by the topological charge (\ref{toc}) is
\begin{equation}\label{tpc}
|Q(\phi(\infty))-Q(\phi(-\infty))|=\frac{4}{3}\sqrt{\frac{\lambda}{2}}v^{3}.
\end{equation}

Now let us consider corresponding NCFT of an NC scalar field ${\hat \phi}$.
We easily read its action as
\begin{equation}\label{nac}
S_{{\rm NC}2}=\int dtdx\left[-\frac{1}{2}\partial_{\mu}{\hat \phi}
\,\partial^{\mu}{\hat \phi}+\frac{v^{2}}{2}{\hat \phi}^{2}
-\frac{\lambda}{4}({\hat \phi}\ast{\hat \phi})^{2}\right],
\end{equation}
where star product between two NC fields, $f$ and $g$, is defined by
\begin{equation}
f(x)\ast g(x)\equiv e^{\frac{i}{2}\theta^{\mu\nu}
\frac{\partial}{\partial\xi^{\mu}}\frac{\partial}{\partial\zeta^{\nu}}}
f(x+\xi )g(x+\zeta)|_{\xi=\zeta=0}.
\end{equation}
In Eq.~(\ref{nac}) we have used the property that one star product can
be replaced by an ordinary product in the action due to spacetime
volume integration. In this simplest NCFT, star products in the
quartic interaction term are the only difference between the
ordinary action (\ref{sac}) and NCFT action (\ref{nac}). In general
this noncommutativity, a way of introducing nonlocality, plays
a key role for complex
but attractive phenomena. At tree level codimension-two
solitons are supported, so-called GMS
solitons~\cite{Gopakumar:2000zd}, and at loop level UV/IR-mixing
was found~\cite{Minwalla:1999px}.
We are interested in codimension-one extended objects given by classical
soliton solutions of Euler-Lagrange equations.
One can easily notice that, in the NC action (\ref{nac}), noncommutativity
appears only in the form of ${\hat \phi}\ast{\hat \phi}$. For any static
object with ${\hat \phi}={\hat \phi}(x)$, every star product is simply
replaced by ordinary product in the NC action (\ref{nac})
\begin{equation}\label{ccc}
\left. {\hat \phi}(x)\ast{\hat \phi}(x)
=e^{\frac{i}{2}\theta^{01}(\partial_{t}^{\xi}\partial^{\zeta}_{y}-
\partial_{x}^{\xi}\partial^{\zeta}_{y})}{\hat \phi}(x+\xi)
{\hat \phi}(x+\zeta)\right|_{\xi=\zeta=0}
={\hat \phi}^{2}(x),
\end{equation}
and so does in equations of motion for the NC scalar field.
Therefore, with identification of ${\hat \phi}=\phi$, the
procedure performed in ordinary scalar field theory can be
repeated in the exactly same way ~\cite{Mukhopadhyay:2002en}.

A point-like kink on a linear sample can be understood
as a straight stringy object
on a planar sample, a flat domain wall in a three-dimensional bulk, and
so on. Mathematically it is nothing but introduction of flat transverse
directions of which coordinates are expressed by $y^{a}$'s $(a=1,2,...)$.
As a first extension let us consider one transverse direction $y$ so that
our spacetime is (2+1)-dimensional, depicted by $(t,x,y)$. A straight
stringy object of our interest requires again only $x$-dependence of
the NC scalar field as ${\hat \phi}={\hat \phi}(x)$.
Though two more exponential type derivative terms between two NC fields,
\begin{equation}
e^{\frac{i}{2}\theta_{ty}\partial_{t}\partial_{y}}~~{\rm and}~~
e^{\frac{i}{2}\theta_{xy}\partial_{x}\partial_{y}},
\end{equation}
appear multiplicatively in ${\hat \phi}\ast{\hat \phi}$, each
exponential becomes trivially unity due to the assumption for the
static and straight kink $\partial_{t}{\hat \phi}(x)
=\partial_{y}{\hat \phi}(x)=0$. It means that addition of one flat
transverse direction does not affect the replacement procedure
from NCFT to ordinary field theory, ${\hat \phi}(x)\ast{\hat
\phi}(x)={\hat \phi}^{2}(x)$. Any generalization to this direction
is trivial, i.e., inclusion of more flat transverse directions
expressed by $(y_{1},y_{2},...)$ does not change the replacement
${\hat \phi}(x)\ast{\hat \phi}(x)\Rightarrow {\hat \phi}^{2}(x)$
in Eq.~(\ref{ccc}). For the kinks, extension to higher polynomial
interaction term ${\hat \phi}\ast{\hat \phi}\ast{\hat \phi}\cdots
\ast{\hat \phi} \Rightarrow {\hat \phi}^{n}$ is also trivial in
any spatial dimensions as far as our static domain wall is flat.

Let us look into a possibility to identify the obtained codimension-one
BPS kink with a stable D$(p-1)$-brane.
Since tension of the D$(p-1)$-brane is given by the energy
(or equivalently the topological charge) of
the kink (\ref{tpc}), ${\cal T}_{p-1}=|Q(\phi(\infty))-Q(\phi(-\infty))|$,
and constant vacuum energy of the action (\ref{sac}) at the unstable
vacuum defines tension of the unstable D$p$-brane
\begin{equation}\label{tns}
{\cal T}_{p}=V(\phi=0)=\frac{\lambda}{4}v^{4},
\end{equation}
one can read a sort of decent relation
\begin{equation}\label{der}
{\cal T}_{p-1}=\frac{8\sqrt{2}}{3}\frac{1}{m_{{\rm H}}}{\cal T}_{p}.
\end{equation}
Since the correct decent relation~\cite{Sen:1999mh} is read as
${\cal T}_{p-1}=\pi R {\cal T}_{p}$ with $R=\sqrt{2}$ for
superstring theory in our unit and
the square of tachyonic pion mass $m_{\phi}^{2}$ near the symmetric vacuum
$\phi=0$ is $m_{\phi}^{2}=-m_{{\rm H}}^{2}=-1/R^{2}$, the obtained
decent relation (\ref{der}) involves about 20$\%$ difference in
its coefficient, i.e., $(\frac{8\sqrt{2}}{3})/\pi=1.20$.

Let us conclude this section by introducing decent relations obtained from
a few effective field theory models of tachyon dynamics for unstable
bosonic D-branes. First example is $\ell =\infty$ field theory model
with nonpolynomial potential
\begin{equation}
S={\cal T}_{p}\int d^{p+1}x\left(-4e^{-\frac{\phi^{2}}{2}}
\partial_{\mu}\phi\partial^{\mu}\phi
-e^{-\frac{\phi^{2}}{2}}\right).
\end{equation}
Note that this tachyon effective action was proposed in Ref.~\cite{Sen:1999md}
and can be understood as two-derivative truncation of
BSFT~\cite{Gerasimov:2000zp}.
Its NC version was also considered in Ref.~\cite{Gopakumar:2000rw}
with NC U(1) gauge field.
By estimating mass of a tachyon kink connecting $T=-\infty$ and $T=\infty$,
we obtain a decent relation
${\cal T}_{p-1}=(2/\sqrt{\pi})\pi R {\cal T}_{p}$~\cite{Minahan:2000tf}.
Since string theory result requires unity instead of $2/\sqrt{\pi}$, it
involves $12.8\%$ difference.
Second example is
$d$-dimensional $p$-adic string theory described by a nonlocal action
\begin{equation}
S=\frac{1}{G_{{\rm s}}^{2}}\frac{p^{2}}{p-1}\int d^{d}x\,
\left(-\frac{1}{2}\phi p^{-\frac{1}{2}\partial^{2}}\phi
+\frac{1}{p+1}\phi^{p+1}\right),
\end{equation}
where $G_{{\rm s}}$ is open string coupling constant and
$p$ is an arbitrary prime
number which is related with tension of the $p$-adic string
$(2\pi\alpha_{p}^{'})^{-1}=(\ln p)^{-1}$.
According to Ref.~\cite{Ghoshal:2000dd}, decent relation between $q$-brane
and $(q-1)$-brane is read as
\begin{equation}
{\cal T}_{q-1}=\sqrt{\frac{2\pi p^{\frac{2p}{p-1}}\ln
p}{p^{2}-1}}\, {\cal T}_{q}\stackrel{p\rightarrow
\infty}{\longrightarrow} \sqrt{2\pi \ln p}\, {\cal T}_{q}= 2\pi
\sqrt{\alpha_{p}^{'}}\, {\cal T}_{q}
\end{equation}
which reproduces an exact decent relation in bosonic string theory
for self-dual radius $R_{{\rm sd}}\stackrel{p\rightarrow\infty}{
=} \sqrt{\alpha_{p}^{'}}$.

In this section we have shown that flat codimension-one static
configurations represented by kinks in ordinary field theory of a
real scalar field are identified by the NC kink solutions of NCFT of a real NC
scalar field.

\setcounter{equation}{0}
\section{Tachyon Kinks in NC Tachyon Field Theory}
Single flat unstable D$p$-brane exists in string theory where
$p$ is odd for type IIA string theory and even for IIB.
Its instability is represented by a real tachyonic degree,
and effective action of the tachyon field $T$ is claimed to be
DBI type~\cite{Garousi:2000tr}
\begin{equation}\label{fa}
S= -\frac{1}{g_{{{\rm s}}}(2\pi)^{\frac{p-1}{2}} } \int d^{p+1}x\;
V(T) \sqrt{-\det (\eta_{\mu\nu} + B_{\mu\nu}+\partial_\mu
T\partial_\nu T)}\, ,
\end{equation}
where flat Minkowski metric $\eta_{\mu\nu}$ is the closed string metric
$g_{\mu\nu}$ of our interest,
$B_{\mu\nu}$ is NS-NS two-form field assumed to be constant,
and string coupling $g_{{\rm s}}$
is inversely proportional to tension of the D$p$-brane as
\begin{equation}\label{ten}
{\cal T}_{p}=\frac{1}{g_{{\rm s}}(2\pi)^{\frac{p-1}{2}}}.
\end{equation}
The tachyon potential $V(T)$ measures variable tension of the unstable
D-brane, i.e., it can be any runaway potential connecting monotonically
its maximum coinciding with the tension of D$p$-brane
and its minimum representing vanishing unstable D$p$-brane.
In our conventions, the potential obeys~\cite{Sen:1999mh}
\begin{equation}\label{vbd}
V(T=0)=1~~\mbox{and}~~ V(T=\pm\infty)=0.
\end{equation}

When $B_{\mu\nu}$ is constant, it is formally
equivalent
with constant DBI type electromagnetic field strength
$F_{\mu\nu}$ on the D$p$-brane.
Since all the tachyon kink configurations supported by the aforementioned
DBI type effective tachyon action (\ref{fa}) are obtained in the constant
DBI type electromagnetic field~\cite{Kim:2003in,Kim:2003ma},
one can just read all the tachyon kinks
in the background of constant NS-NS two-form field $B_{\mu\nu}$ by replacing
each $F_{\mu\nu}$ component by that of $B_{\mu\nu}$.

If we are interested in the kink configurations of codimension-one
$T=T(x)$ with constant background $B_{\mu\nu}$, the action
(\ref{fa}) is then simplified as
\begin{equation}\label{siac}
S=-{\cal T}_{p}\int d^{p+1}x\; V(T)\sqrt{\beta_{p}-\alpha_{px}{T'}^{2}},
\end{equation}
where $\alpha_{px}$ is $11$-component of the cofactor $C^{\mu\nu}$ of matrix
$(X)_{\mu\nu}=\eta_{\mu\nu}+ B_{\mu\nu} + \partial_\mu T\partial_\nu T$
and $\beta_{p}=-\det (\eta_{\mu\nu}+B_{\mu\nu})$.
Rescaling the $x$-coordinate,
\begin{equation}\label{tim}
\tilde{x}=\frac{x}{\sqrt{|\alpha_{px}/\beta_{p}|}},
\end{equation}
the action (\ref{siac}) becomes
\begin{equation}\label{ppab}
S=-{\cal T}_{p}\sqrt{|\alpha_{px}|}
\int dt\int d\tilde{x}\int dy_{1}\cdots dy_{p-1}\,
V(T)\sqrt{\beta_{p}/|\beta_{p}|-(\alpha_{px}/|\alpha_{px}|)\tilde{T}'^{2}},
\end{equation}
where $\tilde{T}'=dT/d\tilde{x}$. Equation of motion is given by
the conservation of $x$-component of pressure $T^{11}=\alpha_{px}
({\cal T}_{p}V/\sqrt{-\det (X)_{\mu\nu}}\,)$, and it is rewritten
by a first-order equation
\begin{equation}\label{hoe}
{\cal E}_{p}=\frac{1}{2}T'^{2}+U_{p}(T),
\end{equation}
where ${\cal E}_{p}=\beta_{p}/2\alpha_{px}$ and $U_{p}(T)=
\alpha_{px}[{\cal T}_{p}V(T)]^{2}/2(T^{11})^{2}$.
Since the equation of motion (\ref{hoe}) has three parameters
$(\beta_{p},\alpha_{px},T^{11})$, all the obtainable kink configurations
are classified by those.

The signature of $T^{11}$ is the same as that of $\alpha_{px}$
so signature of $U_{p}$ is also equal to $\alpha_{px}$.
When $\alpha_{px}$ is negative,
$U_{p}$ becomes upside down and
has minimum value $\alpha_{px}{\cal T}_{p}^{2}/2(T^{11})^{2}$
at $T=0$ and maximum value $0$ at $T=\pm\infty$ due to the runaway nature of
the tachyon potential (\ref{vbd}).
Note that, for $p=1$,
$\alpha_{px}$ should always be nonnegative. As $\beta_{p}$ decreases,
we have (i) constant solution
$T=0$ for $\beta_{p}=\alpha_{px}^{2}{\cal T}_{p}^{2}/(T^{11})^{2}$,
(ii) oscillating solution between $-\pi R \sqrt{-\alpha_{px}/\beta_{p}}
$ and $\pi R \sqrt{-\alpha_{px}/\beta_{p}}$ for
$0<\beta_{p}<\alpha_{px}^{2}{\cal T}_{p}^{2}/(T^{11})^{2}$,
(iii) monotonic increasing (decreasing) solution connecting $T=-\infty$
and $T=\infty$ with ${\tilde T}^{'}(\pm\infty)=0$ for $\beta_{p}=0$, and
(iv) another monotonic increasing (decreasing) solution connecting $T=-\infty$,
and $T=\infty$ with ${\tilde T}^{'}(\pm\infty)\ne 0$ for $\beta_{p}<0$.
When $\alpha_{px}$ is positive,
$U_{p}$ has maximum value
$\alpha_{px}{\cal T}_{p}^{2}/2(T^{11})^{2}$
at $T=0$ and minimum value $0$ at $T=\pm\infty$.
As $\beta_{p}$ increases, we have (v) constant solution
$T=\pm\infty$ for $\beta_{p}=0$, (vi) convex-down (-up) solution connecting
its minimum (maximum) and positive (negative) infinity for
$0<\beta_{p}<\alpha_{px}^{2}{\cal T}_{p}^{2}/(T^{11})^{2}$,
(vii) monotonic increasing
(decreasing) solution connecting $T=0$ and $T=\infty~(-\infty)$ with
(viii) constant $T=0$ solution for
$\beta_{p}=\alpha_{px}^{2}{\cal T}_{p}^{2}/(T^{11})^{2}$,
and (ix) another monotonic increasing
(decreasing) solution connecting $T=-\infty~(\infty)$ and $T=\infty~(-\infty)$.
Except unstable and stable vacuum solutions at $T=0$ and $T=\pm\infty$,
the obtained solutions are interpreted as array of kink-antikink for (ii),
topological kinks for (iii) and (iv), bounce for (vi), half-kink for
(vii), and hybrid of two half-kinks for (ix). When $\alpha_{px}$=0,
there exists no nontrivial solution.

Here let us assume a specific tachyon potential satisfying
Eq.~(\ref{vbd})~\cite{Buchel:2002tj,Kim:2003he,Leblond:2003db}
\begin{equation}\label{V3}
V(T)=\frac{1}{\cosh \left(\frac{T}{R}\right)},
\end{equation}
where $R$ is $\sqrt{2}$ for the non-BPS D-brane in the superstring
and 2 for the bosonic string.
This form of the potential has been derived in open string theory
by taking into account
the fluctuations around $\frac{1}{2}$S-brane configuration
with the higher derivatives neglected, i.e., $\partial^2 T = \partial^3 T=
\cdots = 0$~\cite{Kutasov:2003er}.

For the $1/\cosh$-potential (\ref{V3}) the tachyon equation (\ref{hoe})
is solved and we obtain exact kink
solutions~\cite{Kim:2003in,Kim:2003ma},
\begin{equation}\label{rss3}
\sinh\left(\frac{T(x)}{R}\right)=
\left\{
\begin{array}{ccllc}
\sqrt{u^{2}+1}\sinh \left(x/\zeta\right)
&\mbox{for}& \alpha_{px}<0, & \beta_{p}<0 & (\mbox{iv})\\
ux/\zeta
&\mbox{for}&\alpha_{px}<0, & \beta_{p}=0 & (\mbox{iii})\\
\sqrt{u^{2}-1} \sin\left(x/\zeta\right)
&\mbox{for}&\alpha_{px}<0, &
0<\beta_{p}<\alpha_{px}^{2}{\cal T}_{p}^{2}/(T^{11})^{2}& (\mbox{ii})\\
\sqrt{u^{2}-1} \cosh\left(x/\zeta\right)
&\mbox{for}&\alpha_{px}>0, & 0<\beta_{p}
<\alpha_{px}^{2}{\cal T}_{p}^{2}/(T^{11})^{2} & (\mbox{vi})\\
\exp \left(x/\zeta\right)
&\mbox{for}&\alpha_{px}>0, & \beta_{p}=
\alpha_{px}^{2}{\cal T}_{p}^{2}/(T^{11})^{2} & (\mbox{vii})\\
\sqrt{1-u^2} \sinh\left(x/\zeta\right)
&\mbox{for}&\alpha_{px}>0, & \beta_{p}>
\alpha_{px}^{2}{\cal T}_{p}^{2}/(T^{11})^{2} & (\mbox{ix})
\end{array}
\right. ,
\end{equation}
where the scales in the solutions are
\begin{equation}
u^{2}=\left|\frac{{\cal T}_{p}^{2}\alpha_{px}^{2}}
{\beta_{p}(T^{11})^{2}}\right|,
\quad \zeta=R\sqrt{\left| \frac{\alpha_{px}}{\beta_{p}}\right|}\, .
\end{equation}
Computation of tension and F1 charge leads to
interpretation of D$(p-1)$-brane (and D$(p-1)$F1) for single topological kink
in (iii) and (iv), and array of D$(p-1)\bar{{\rm D}}(p-1)$
(and D$(p-1)$F1-$\bar{{\rm D}}(p-1)$F1) for array of kink-antikink in (ii),
where F1 is confined on D$(p-1)$ in the form of string fluid.
For (ii) and (iii), BCFT calculation confirms this
interpretation~\cite{Sen:2003bc}. The functional form of
last three solutions in Eq.~(\ref{rss3}) coincide with exact rolling tachyon
solutions~\cite{Kim:2003he,Kim:2004zq}. This can be explained by the signatures
in the rescaled action (\ref{ppab}), i.e., $\beta_{p}/|\beta_{p}|=1$
and $\alpha_{px}/|\alpha_{px}|=1$ for those solutions, which coincide with
those in the action for the rolling tachyons.

In relation with BPS nature, single topological kink (iii) saturates
BPS type bound with thickness for $T^{11}\ne 0$ but each kink (antikink)
in the array (iv) becomes a BPS object only when its thickness vanishes
($T^{11}\rightarrow 0$). In what follows let us study codimension-one
tachyon solitons in NC DBI type action.

From the action (\ref{fa}), the flat closed string metric $\eta_{\mu\nu}$
and the NS-NS two-form field $B_{\mu\nu}$ are replaced by
open string metric $G_{\mu\nu}$ and noncommutativity parameter
$\theta^{\mu\nu}$
\begin{eqnarray}
G_{\mu\nu}&=&\eta_{\mu\nu}-(B\eta^{-1}B)_{\mu\nu},
\label{omet}\\
\theta^{\mu\nu}&=&-\left(\frac{1}{\eta+B}B\frac{1}{\eta-B}\right)^{\mu\nu},
\label{ncpa}
\end{eqnarray}
where $-\det G_{\mu\nu}\ge 0$. Once the tachyon field $T$ is
turned on in the effective action (\ref{fa}), corresponding NC
tachyon ${\hat T}$ should be introduced in the NC action. In
Ref.~\cite{Banerjee:2004cw}, a candidate was proposed
\begin{equation}\label{nct}
\hat{S}= -\frac{{\hat {\cal T}}_p}{2} \int d^{p+1}x\;
\left[\hat{V}(\hat{T}) \ast \sqrt{ -\det {}_{\ast}\left[G_{\mu\nu}+
\frac{1}{2}(\partial_{\mu}\hat{T}\ast\partial_{\nu}\hat{T}
+\partial_{\nu}\hat{T}\ast\partial_{\mu}\hat{T})\right]}
+(\sqrt{\hspace{-5mm}}\leftrightarrow \hat{V}) \right],
\end{equation}
where comparison of two actions (\ref{fa}) and (\ref{nct}) with turning off
both tachyon fields leads to a relation between the tensions
\begin{equation}\label{ret}
{\hat {\cal T}}_p\equiv \frac{1}{G_{{{\rm s}}}(2\pi)^{\frac{p-1}{2}}}
=(-G)^{-1/4}{\cal T}_{p}=\frac{{\cal T}_{p}}{\sqrt{\det(1+\eta^{-1}B)}},
\end{equation}
where $G=\det(G_{\mu\nu})$.
The determinant with subscript $\ast$ in the action (\ref{nct}) denotes
\begin{eqnarray}\label{ncdet01}
{\det}_\ast \hat X_{\mu\nu} = \frac{1}{(p+1)!}
\epsilon_{\mu_1\mu_2\cdots \mu_{p+1}} \epsilon_{\nu_1\nu_2\cdots
\nu_{p+1}} \hat X_{\mu_1\nu_1}\ast \hat X_{\mu_2\nu_2}\ast \cdots
\ast \hat X_{\mu_{p+1}\nu_{p+1}}.
\end{eqnarray}
The NC tachyon potential ${\hat V}({\hat T})$ imitates properties of
that in EFT, so that in general ${\hat V}({\hat T})$ can have any functional
form satisfying ${\hat V}({\hat T}=0)=1$ and ${\hat V}({\hat T}=\pm\infty)=0$
as given in Eq.~(\ref{vbd}) and the specific form of our interest is
again $1/{\rm cosh}$ form:
\begin{eqnarray}
\label{ncpot}
\hat{V} ({\hat T}) &\equiv &  1 - \frac{1}{2} \frac{\hat T}{R}\ast
\frac{\hat T}{R} + \frac{5}{24}\frac{\hat T}{R}\ast
\frac{\hat T}{R}\ast \frac{\hat T}{R}\ast \frac{\hat T}{R} +\, \cdots
\nonumber \\
&=& 1 + \sum^{\infty}_{k=1} \frac{E_{2k}}{(2k)!} \left[
\left(\frac{\hat T}{R}\right)^{2k}\right]_\ast
= \left[ \frac{1}{\cosh ({\hat T}/R)}\right]_\ast,
\end{eqnarray}
where $\left[\quad \right]_\ast$ stands for
\begin{eqnarray}
\left[\hat A_1 \hat A_2 \cdots \hat A_n\right]_\ast &=&
\frac{1}{n!} \left(\hat A_1 \ast \hat A_2 \ast \cdots \ast \hat
A_n + \hat A_1 \ast \hat A_3  \ast \cdots \ast \hat A_n\right.
\nonumber\\
&& \left.\qquad + \cdots
(\mbox{all~possible~permutations})\right.\Big).
\label{stnot}
\end{eqnarray}

Since the objects of our interest are flat static
kink solutions given with a function
of a coordinate $x$ as
\begin{equation}\label{asta}
{\hat T}={\hat T}(x)
\end{equation}
in the flat background
with constant $G_{\mu\nu}$ and $\theta^{\mu\nu}$, every star product of two
NC fields is always replaced by ordinary multiplication as
\begin{equation}\label{mul}
{\hat T}\ast{\hat T}={\hat T}^{2},\quad
(\partial_{\mu}{\hat T})\ast{\hat T}=\delta_{\mu 1}{\hat T}'{\hat T},\quad
\partial_{\mu}{\hat T}\ast\partial_{\nu}{\hat T}=
\delta_{\mu 1}\delta_{\nu 1}{\hat T}'^{2}.
\end{equation}

Substituting the formulae (\ref{mul}) into the action (\ref{nct}), we obtain
again a simplified action
\begin{equation}\label{sia2}
{\hat S}=-{\hat {\cal T}}_{p}\int d^{p+1}x\, {\hat V}({\hat T})
\sqrt{{\hat \beta}_{p}-{\hat \alpha}_{px}{\hat T}'^{2}}\,,
\end{equation}
where ${\hat \beta}_{p}=-\det (G_{\mu\nu})\ge 0$ and
${\hat \alpha}_{px}$ is $11$-component of the cofactor
${\hat C}^{\mu\nu}$
of matrix $({\hat X})_{\mu\nu}=G_{\mu\nu}+ \partial_\mu {\hat T}
\partial_\nu {\hat T}$.
When $\hat\beta_p$ is not zero, we can rewrite the action (\ref{sia2}) as
\begin{equation}\label{sia3}
{\hat S} = -\hat{{\cal T}}_{p}\sqrt{|\hat\alpha_{px}|}
\int dt\int d\tilde{x}\int dy_{1}\cdots dy_{p-1}\,
\hat V(\hat T)\sqrt{1-\frac{\hat\alpha_{px}}{|\hat\alpha_{px}|}
\left(\frac{d\hat T}{d\tilde x}\right)^2}\,,
\end{equation}
where
\begin{eqnarray}
\tilde x \equiv \sqrt{\frac{\hat\beta_p}{|\hat\alpha_{px}|}}\, x.
\nonumber
\end{eqnarray}

Note that the action (\ref{sia2}) is formally the same as that in
Eq.~(\ref{siac}) with replacing the quantities without hat by those
with hat, e.g.,
\begin{equation}\label{repl}
{\cal T}_{p}\leftrightarrow {\hat {\cal T}}_{p},
\quad T\leftrightarrow {\hat T},\quad V \leftrightarrow {\hat V},\quad
{\beta}_{p}\leftrightarrow {\hat \beta}_{p},\quad {\alpha}_{px}
\leftrightarrow {\hat \alpha}_{px}.
\end{equation}
Therefore, we can automatically read the equation of motion for
static codimension-one objects as
\begin{equation}\label{hhoe}
{\hat {\cal E}}_{p}=\frac{1}{2}{\hat T}'^{2}+{\hat U}_{p}({\hat T}),
\end{equation}
where
\begin{equation}\label{hho1}
{\hat {\cal E}}_{p}=\frac{{\hat \beta}_{p}}{2{\hat \alpha}_{px}},\qquad
{\hat U}_{p}({\hat T})= \frac{\hat\alpha_{px}}
{2 \hat\beta_p({\hat T}^{11})^{2}}
[{\hat {\cal T}}_{p}{\hat V}({\hat T})]^{2}.
\end{equation}
Here we must caution that $ (T^{11})^{2}$ is replaced by $({\hat
T}^{11})^{2}\hat\beta_p$. This is just because that the
NC energy-momentum tensor in NCFT is defined under the open string
metric $G_{\mu\nu}$:
\begin{eqnarray}
\hat{T}^{\mu\nu}&\equiv &  \frac{2}{\sqrt{-G}}
\frac{\delta\hat{S}}{\delta G_{\mu\nu}}
\label{emte} \\
&=& \frac{\hat {\cal T}_p \hat V
\hat C^{\mu\nu}_{{\rm S}}}{ \sqrt{-G}
\sqrt{-\hat X}}
\label{emco} \\
&=&\frac{\hat {\cal T}_p \hat V }{\sqrt{-\hat X} \sqrt{-G}}\left(
\hat X G^{\mu\nu} - G G^{\mu\lambda} \partial_\lambda \hat T G^{\nu\kappa}
\partial_\kappa \hat T\right),
\label{emex}
\end{eqnarray}
where $\hat X \equiv\det\hat X_{\mu\nu}$, ${\hat C}^{\mu\nu}_{{\rm
S}}$ is symmetric part of the cofactor ${\hat C}^{\mu\nu}$, and
$x$-component of conservation of NC energy-momentum tensor
\begin{eqnarray}\label{ncc}
{\partial}_{\mu} \hat{T}^{\mu\nu}=0
\end{eqnarray}
forces ${\hat T}^{11}$ to be a constant.

In order to compare classical NC tachyon kink solutions
which will be obtained in the subsections in 3.1--3.6 and
those in BCFT, we should compare components of energy-momentum tensor
and string current density, i.e., they are $T_{\mu\nu}$ vs.
$T_{\mu\nu}^{{\rm BCFT}}$ and $J_{\mu\nu}$ vs. $J^{{\rm BCFT}}_{\mu\nu}$.
To be specific, the energy-momentum tensor $T_{\mu\nu}$ and the
string current density $J_{\mu\nu}$ are given by responses to small
fluctuations of the background closed string metric $g_{\mu\nu}$
and the anti-symmetric tensor field $B_{\mu\nu}$:
\begin{eqnarray}
T_{\mu\nu}&\equiv& \left. -\frac{2}{\sqrt{-g}}\frac{\delta \hat S}{\delta
g^{\mu\nu}}\right|_{g_{\mu\nu}=\eta_{\mu\nu}} \nonumber \\
&=&\left.
\frac{{\cal T}_p\,  V}{
(gG)^{1/4} \sqrt{-\hat X}}
\left[\frac{1}{2}\hat X g_{\mu\nu} +(g_{\mu\lambda} g_{\nu\kappa} -
B_{\mu\lambda} B_{\nu\kappa})\left(\frac{1}{2}\hat X G^{\lambda\kappa}
-G\partial^\lambda\hat T \partial^\kappa\hat T\right)\right]
\right|_{g_{\mu\nu}=\eta_{\mu\nu}} ,
\nonumber \\
\label{emt1} \\
J^{\mu\nu} &\equiv& \left. \frac{2}{\sqrt{-g}}\frac{\delta \hat S}{
\delta B_{\mu\nu}} \right|_{g_{\mu\nu}=\eta_{\mu\nu}} \nonumber \\
&=& \left. -\frac{{\cal T}_p\,  V}{ (gG)^{1/4} \sqrt{-\hat X}}
\left[\left\{-\frac{1}{2} \hat X
G^{\mu\lambda}B_{\lambda\kappa}g^{\kappa\nu} + G\partial^\mu\hat
T\partial^\lambda \hat T B_{\lambda\kappa}g^{\kappa\nu}\right\} -
\left\{\mu\,\, \leftrightarrow \,\, \nu\right\} \right]
\right|_{g_{\mu\nu}=\eta_{\mu\nu}} .
\nonumber \\
\label{KR1}
\end{eqnarray}
For the derivation of Eqs.~(\ref{emt1})--(\ref{KR1}), we used
\begin{eqnarray}
\delta G_{\mu\nu} &=& -\left(g_{\mu\lambda} g_{\nu\kappa}
-B_{\mu\lambda} B_{\nu\kappa}\right) \delta g^{\lambda\kappa}
\nonumber \\
&& -\frac{1}{2} \left(\delta_\mu^{~~\lambda} g^{\kappa\rho}
B_{\rho\nu} + \delta_\nu^{~~\lambda} g^{\kappa\rho} B_{\rho\mu}
- (\lambda \,\, \leftrightarrow \,\,\kappa) \right)
\delta B_{\lambda\kappa}, \nonumber
\end{eqnarray}
and note that Eqs.~(\ref{emt1})--(\ref{KR1})
coincide with those from commutative tachyon
effective action~\cite{Kim:2003in, Kim:2003ma}.

The equation (\ref{hhoe}) derived
from the simplified action (\ref{sia2}) is consistent with
the NC tachyon equation derived
directly from the original NC action (\ref{nct}).
First observation is made by comparing the expressions in Eq.~(\ref{hhoe})
with those in Eq.~(\ref{hoe}): One is
\begin{equation}\label{etoe}
{\hat {\cal E}}_{p}=\frac{{\hat \beta}_{p}}{2{\hat \alpha}_{px}}
=\frac{{\beta}_{p}}{2{\alpha}_{px}}={\cal E}_{p}
\end{equation}
with the help of Eq.~(\ref{omet}), and, the other is, when the NC
tachyon field $\hat{T}(x)$ is identified by the tachyon field
$T(x)$ of EFT which results in ${\hat V}({\hat T})=V(T)$,
\begin{equation}\label{uteq}
 {\hat U}_{p}= U_{p}
\end{equation}
with the help of Eq.~(\ref{ret}).
From Eq.~(\ref{emco}), we read the relation between ${\hat T}^{11}$ and
$T^{11}$
\begin{equation}\label{txr}
\sqrt{-G}\, {\hat T}^{11}=T^{11},
\end{equation}
where we have used ${\hat C}^{11}={\hat C}^{11}|_{{\hat T}=0}$
and $\sqrt{-X/-{\hat X}}=\sqrt{-\det (\eta+B)
/-\det (G_{\mu\nu})}$.
It means that every kink solution obtained from the EFT
given in the first half of this section can also be a solution of the
corresponding NCFT of the NC tachyon (\ref{hhoe}) as far as
the aforementioned relations between Eq.~(\ref{etoe}) and Eq.~(\ref{uteq})
are satisfied.

We do not need to recapitulate presenting species of the solutions
here due to the aforementioned analysis between Eq.~(\ref{hoe}) and
Eq.~(\ref{rss3}). Instead we summarize the viable NC kink solutions in Table~1,
specified by the value and signature of ${\hat {\cal E}}_{p}$ and
${\hat U}_{p}({\hat T}=0)$.
In particular,
${\hat U}_{p}({\hat T}=0)$ should be nonpositive for
$p=1$. Therefore, only the first three types of NC kinks in Table~1
are obtained from the unstable D1-brane~\cite{Banerjee:2004cw}.
On the other hand, ${\hat U}_{p}({\hat T}=0)$ can be arbitrary for $p\ge 2$
so that all the six types of NC kinks in Table~1 can be obtained.

If we naively look at the expressions of ${\hat {\cal E}}_{p}$ and
${\hat U}_{p}$ in Eq.~(\ref{hho1}), signature of ${\hat {\cal
E}}_{p}$ and that of ${\hat U}_{p}$ should be the same and equal
to that of ${\hat {\alpha}}_{px}$ due to nonnegativity of ${\hat
\beta}_{p}$, i.e., ${\hat
\beta}_{p}=-G=-\det(G_{\mu\nu})=[-\det(\eta+B)]^{2}\ge 0$. If it
is indeed the case, then the topological NC kink (iv) in the
second low of Table~1 cannot be supported. This observation was
made under the assumption that all the squared quantities in
Eq.~(\ref{hho1}) are nonnegative. This needs not to be true for
${\hat {\cal T}}_{p}$ introduced in Eq.~(\ref{ret}). The condition
we have is $-G\ge 0$, so imaginary $(-G)^{-1/4}=1/\sqrt{\det
(1+\eta^{-1}B)}$ is not excluded. For this case, the tension
${\cal T}_{p}$ is real positive and thereby ${\hat {\cal T}}_{p}$
(\ref{ret}) is imaginary. This phenomenon corresponds to
electromagnetic fields larger than critical value with
$-\det(\eta+B)<0$ in the EFT, and this situation is not allowed
without the NC tachyon $({\hat T}=0)$ since the NC action
(\ref{sia2}) becomes imaginary. Once we turn on the NC tachyon for
a positive ${\hat \alpha}_{px}$ corresponding to the case (iv), of
our interest, the square root part of the NC action (\ref{sia2})
provides additional imaginary number as follows
\begin{equation}\label{abt}
\sqrt{{\hat \beta}_{p}-{\hat \alpha}_{px}{\hat T}'^{2}}
=i\sqrt{|{\hat \alpha}_{px}|{\hat T}'^{2}-{\hat \beta}_{p}},
\end{equation}
and conclusively
the NC action proportional to ${\hat {\cal T}}_{p}\sqrt{{\hat \beta}_{p}
-{\hat \alpha}_{px}{\hat T}'^{2}}$ becomes real. Other physical quantities,
e.g., the NC energy-momentum tensor (\ref{emco}), are also real.
Up to the present point, there is no reason to neglect
the topological NC kink solution~(iv).

\begin{center}
\renewcommand{\arraystretch}{1.4}
\begin{tabular}{|c | c | c |c|} \hline \hline
${\hat U}_{p}(0)$ & ${\hat {\cal E}}_{p}$ & Type of NC kinks & Exact solutions
\\ \hline \hline
& ${\hat {\cal E}}_{p}>0$ & Topological NC kink & (iv)\\
${\hat U}_{p}(0)<0$ & ${\hat {\cal E}}_{p}=0$ & Thick topological BPS NC kink
& (iii)\\
& ${\hat U}_{p}(0)<{\hat {\cal E}}_{p}<0$ & Array of NC kink-antikink &
(ii)\\ \hline
& $0<{\hat {\cal E}}_{p}<{\hat U}_{p}(0)$ & NC bounce & (vi)\\
${\hat U}_{p}(0)>0$ & ${\hat {\cal E}}_{p}={\hat U}_{p}(0)$ &
NC half-kink & (vii)\\
& ${\hat {\cal E}}_{p}>{\hat U}_{p}(0)$ & Hybrid of two NC half-kinks &
(ix)\\ \hline\hline
\end{tabular}
\end{center}
\begin{center}{
Table 1: List of regular static NC kink configurations. Functional form of
exact NC kink solutions in the fourth column are read from Eq.~(\ref{rss3})
under the replacement (\ref{repl}).}
\end{center}

From now on, let us take into account the $1/\cosh$ NC tachyon
potential (\ref{ncpot}) and study detailed properties of the NC
kinks, including computation of tension, and their various limits.
In subsections 3.1--3.3, we consider the case of $\hat U_p(0) <
0$. As given in the upper half of Table~1, there exist three types
of NC kink solutions of the equation of motion (\ref{hhoe}). Then,
in subsections 3.4--3.5, we consider the case of $\hat U_p(0) >0$.
As given in the lower half of Table~1, there exist three more
types of extended objects from the equation of motion
(\ref{hhoe}). These three types of solutions cannot be obtained in
the case $p=1$ since the signature of $\hat U_p$ cannot be flipped
due to negativity of $\hat \alpha_{1x}$ for the unstable
D1-brane~\cite{Banerjee:2004cw}.

\subsection{Array of kink and antikink for $\hat U_p(0) < \hat {\cal E}_p <0$}

When $\hat U_p(0) < \hat {\cal E}_p <0$ ($\hat {\cal T}_p^2 >0$
$\&$ $\hat\alpha_{px} <0$), we find an exact solution of Eq.~(\ref{hhoe})
\begin{equation}\label{nsol1}
\sinh\left(\frac{\hat T(x)}{R}\right)
= \pm \sqrt{\hat u^2 -1}\, \sin \left(\frac{x}{{\hat \zeta}}\right),
\end{equation}
where
\begin{equation}\label{uzeta}
\hat u^{2}=\left|\frac{\hat {\cal T}_{p}^{2}\hat\alpha_{px}^{2}}{
\hat\beta_{p}^2({\hat T}^{11})^{2}}\right|,
\quad \hat\zeta=R\sqrt{\left| \frac{\hat\alpha_{px}}{\hat\beta_{p}}\right|}\, .
\end{equation}
The NC tachyon field  $\hat T(x)$ oscillates spatially between
$\hat T_{{\rm max}} = R \cosh^{-1} \hat u$ and $-\hat T_{{\rm
max}}$ with finite period $2\pi\hat\zeta$. Then the obtained
configuration (\ref{nsol1}) is interpreted as an array of NC
kink-antikink. If we consider half period of the configuration,
which comprises single kink or single antikink, NC energy density of
the kink (antikink) provides decent relation of the
codimension-one object
\begin{eqnarray}
\frac{{\hat H}}{\int dy_{1}\cdots dy_{p-1}} &\equiv&
-\int_{-\frac{\pi}{2}\hat\zeta}^{\frac{\pi}{2}\hat\zeta}
dx\, \sqrt{-G}\, \hat T^0_{~~0}
\label{ede} \\
&=& \hat {\cal T}_p \sqrt{\hat\beta_{p}\hat u^2}
\int_{-\frac{\pi}{2}\hat\zeta}^{\frac{\pi}{2}\hat\zeta} dx
\frac{1}{1 + (\hat u^2 -1)\sin^2 (x/\hat\zeta)}
\nonumber \\
&=&\pi R \hat {\cal T}_p \sqrt{-\hat\alpha_{px}}
=\pi R {\cal T}_p\sqrt{-\alpha_{px}}\, ,
\label{arts}
\end{eqnarray}
where we used the formula of NC energy-momentum tensor (\ref{emex})
for ${\hat T}^0_{~~0}$ in Eq.~(\ref{ede}), and Eq.~(\ref{ret}) and
Eqs.~(\ref{etoe})--(\ref{uteq}) for the last line (\ref{arts}).
Since the factor $\sqrt{-\alpha_{px}}$ in Eq.~(\ref{arts}) also
appeared when ${\rm F}1$ is confined on D$(p-1)$-brane in ordinary
EFT~\cite{Kim:2003ma}
\begin{equation}\label{dece}
{\cal T}_{p-1}=\pi R {\cal T}_p \sqrt{-\alpha_{px}}\, ,
\end{equation}
we arrive at a conclusion that ${\hat H}/\int dy_{1}\cdots dy_{p-1}$
is nothing but
the tension of a unit NC kink, i.e., ${\hat H}/\int dy_{1}\cdots dy_{p-1}
={\cal T}_{p-1}=(-G)^{(1/4)}\hat {\cal T}_{p-1}$.

The commutative limit $\theta^{\mu\nu}\rightarrow 0$ corresponds to
vanishing $B_{\mu\nu}$ limit (\ref{ncpa}) and then smoothly continues
to Minkowski spacetime (\ref{omet}). Then
$-\hat\alpha_{px}= \hat\beta_p=-\alpha_{px}= \beta_p=1$ and the decent
relation without DBI type electromagnetic field is reproduced.

In the limit $\hat T^{11} \to 0^-$, boundary value of the tachyon field
$T_{{\rm max}}$ approaches rapidly
a true vacuum at infinity, $\lim_{T_{\rm max}\to\infty} V(T_{\rm max}
)\rightarrow 0$, with keeping the half period
$\pi{\hat \zeta}$ fixed, independent of the value of ${\hat T}^{11}$.
Therefore, each kink (antikink) becomes a topological kink (antikink), and
its NC energy density
$-\sqrt{-G}\,\hat T^0_{~~0}$ is sharply peaked at each localized point
of a kink (antikink). The NC energy density profile
for the configuration (\ref{nsol1}) is expressed by a sum of $\delta$-functions
\begin{equation}\label{arthin}
\lim_{\hat T^{11} \to 0^-} \hat {\cal H}_p =
\lim_{\hat T^{11} \to 0^-}  -\sqrt{-G}\,\hat T^0_{~~0}
= \pi R \hat {\cal T}_p \sqrt{- \hat\alpha_{px}} \sum_{n= -\infty}^{\infty}
\delta \left( x - n\pi \zeta\right),
\end{equation}
so that each topological kink (antikink) in the array is a BPS-like
object~\cite{Sen:2003zf}.
This singular limit of unit kink (antikink) with topological and BPS nature
can universally be achieved for every tachyon potential satisfying
Eq.~(\ref{vbd})~\cite{Sen:2003tm}.

In case of the array of kink and antikink, BCFT solution is already known
in the presence of constant electric field~\cite{Sen:2003bc}.
Here let us compare the BCFT result and the NCFT solution.
By inserting the solution (\ref{nsol1}) into
the energy-momentum tensor (\ref{emt1}) and the string current density
(\ref{KR1}), we obtain nonvanishing components of them
\begin{eqnarray}
T_{00} &=& \frac{{\cal T}_p \hat V}{\sqrt{G_0 + {\hat T}'^2}}
\left(1 + \hat{T}'^2\right)
\label{eT00}\\
&=&\frac{{\cal T}_p E_0^2}{\sqrt{G_0}}\cos^2 (\pi\tilde\lambda)
+ {\cal T}_p \sqrt{G_0}\, \tilde f(\sqrt{G_0} x/R),
\nonumber \\
T_{11} &=& -\frac{{\cal T}_p \hat V}{\sqrt{G_0 + {\hat T}'^2}}
= -\frac{{\cal T}_p}{ \sqrt{G_0}}\cos^2 (\pi\tilde\lambda),
\label{eT11} \\
T_{ab} &=& - {\cal T}_p \hat V\, \sqrt{G_0 + \hat{T}'^2}\, \delta_{ab}
=- {\cal T}_p \sqrt{G_0}\, \tilde f(\sqrt{G_0} x/R ) \delta_{ab},
\label{eTab} \\
J^{01} &=& \frac{{\cal T}_p \hat V E_0}{\sqrt{G_0 + \hat{T}'^2}}
= \frac{{\cal T}_p E_0}{ \sqrt{G_0}}\cos^2 (\pi\tilde\lambda),
\label{eF1}
\end{eqnarray}
where $\tilde\lambda$ is a parameter labelling the initial
pressure $T_{11}$, $\tilde f(x)$ is given by
\begin{eqnarray}\label{tfx}
\tilde f(\sqrt{G_0} x/R) = \frac{\cos^2(\pi\tilde\lambda)}{
\cos^4(\pi\tilde\lambda) +\left[1 +\cos^2(\pi\tilde\lambda)\right]
\sin^2(\pi\tilde\lambda) \sin^2\left(\sqrt{G_0} x/R\right)},
\end{eqnarray}
and nonvanishing components of the open string metric $G_{\mu\nu}$ and
noncommutative parameter $\theta^{\mu\nu}$ in this background are
given by
\begin{eqnarray}
&&-G_{00}=G_{11}=1-E_{0}^{2}\stackrel{{\rm set}}{=}G_{0},
\quad G_{ab}=\delta_{ab}~(a,b=2,3,\cdots, p),
\\
&&\theta^{01}=-\theta^{10}=\frac{E_{0}}{1-E_{0}^{2}}
\stackrel{{\rm set}}{=}\theta .
\end{eqnarray}
Since the BCFT results obtained through double Wick rotation~\cite{Sen:2003bc}
are for superstring theory
\begin{eqnarray}
T^{{\rm BCFT}}_{00}
&=& \frac{{\cal T}_p E_0^2}{ \sqrt{G_0}}\cos^2 (\pi\tilde\lambda)
+ {\cal T}_p\sqrt{G_0}\, f(\sqrt{G_0}\, x),
\label{bT00} \\
T^{{\rm BCFT}}_{11}
&=& -\frac{{\cal T}_p}{ \sqrt{G_0}}\cos^2 (\pi\tilde\lambda),
\label{bT11} \\
T^{{\rm BCFT}}_{ab} &=& - {\cal T}_p\sqrt{G_0}\, f(\sqrt{G_0}\, x)\delta_{ab},
\label{bTab}\\
J^{01}_{{\rm BCFT}} &=& -T_{11}^{{\rm BCFT}}\, E_0,
\label{bF1}
\end{eqnarray}
where
\begin{eqnarray}\label{fx}
f(\sqrt{G_0}\, x) = \frac{\left[1 + \sin^2(\pi\tilde\lambda)\right] \cos^2
(\pi\tilde\lambda)}{ \cos^4 (\pi\tilde\lambda) + 4 \sin^2
(\pi\tilde\lambda) \sin^2(\sqrt{G_0}\,x/\sqrt{2})}.
\end{eqnarray}
When we identify the pressure component along the transverse direction
$T_{11}=T^{{\rm BCFT}}_{11}$ in Eq.~(\ref{eT11}) and Eq.~(\ref{bT11})
with $R=\sqrt{2}$,
the string charge densities coincide exactly as in Eq.~(\ref{eF1})
and (\ref{bF1}). On the other hand,
the energy densities (\ref{eT00})--(\ref{bT00})
and the other nonvanishing pressure components (\ref{eTab})--(\ref{bTab})
are qualitatively agreed but do not match exactly due to difference between
$\tilde f$ (\ref{tfx}) and $f$ (\ref{fx}). For bosonic string theory
this kind of relations also holds.

\subsection{Topological BPS NC kink for
$\hat U_p(0) < 0$ and $\hat {\cal E}_p =0$}

When $ \hat {\cal E}_p =0$ ($\hat {\cal T}_p^2 >0$ $\&$
$\hat\alpha_{px} <0$),
i.e., $\hat\beta_p = 0$, we find a regular monotonic solution with
boundary conditions $T(\pm\infty)=\pm ({\rm or}~\mp)\infty$
\begin{equation}\label{nsol2}
\sinh\left(\frac{\hat T(x)}{R}\right) = \pm \frac{\hat u}{\hat\zeta} x
=\pm \left|\frac{{\hat {\cal T}}_p \sqrt{-\hat\alpha_{px}}}{R\hat T^{11}
\sqrt{\hat\beta_p}}\right| x,
\end{equation}
and it is single topological kink for $(\pm)$-signature (antikink
for $(\mp)$-signature). Here we must note that the slope
$\left|\frac{{\hat {\cal T}}_p \sqrt{-\hat\alpha_{px}}}{R{\hat T}^{11}
\sqrt{\hat\beta_p}}\right|$ in the Eq.~(\ref{nsol2}) is finite in
the limit $\hat\beta_p \to 0$ because 11-component of NC
energy-momentum tensor is expressed by
\begin{eqnarray}\nonumber
\hat T^{11} = \frac{\hat{\cal T}_p \hat V \hat\alpha_p}{
\sqrt{\hat\beta_p -\hat\alpha_{px} {T'}^2}\sqrt{\beta_p}}.
\end{eqnarray}
The obtained solution (\ref{nsol2}) can also be reproduced from
infinite period limit,
$\lim_{\hat\beta_p \to 0} 2\pi\hat{\zeta} \rightarrow \infty$, at each
kink site of the array (\ref{nsol1}),
i.e., $\sin(x/\hat{\zeta})\sim (-1)^{n+1}\left[(x/\hat{\zeta})-n\pi\right]$
for a given $n$.

Since ${\hat \beta}_{p}=0$, the action (\ref{sia2}) divided by negative
time integral and transverse volume is easily computed
\begin{eqnarray}\label{sia4}
\frac{\hat S}{-\int dt \int dy_{1}\cdots dy_{p-1}}
&=& \pm\hat {\cal T}_p \sqrt{-\hat\alpha_{px}}\,
\int_{-\infty}^{\infty} dx\, \hat V \left(\frac{d\hat T}{dx}\right)
\nonumber \\
&=& \hat {\cal T}_p \sqrt{-\hat\alpha_{px}}\,
\int_{-\infty}^{\infty} d\hat T\, \hat V (\hat T)
\label{topc}\\
&=& \pi R \hat {\cal T}_p \sqrt{-\hat\alpha_{px}}\,,
\label{dere}
\end{eqnarray}
where $+(-)$ in the first line corresponds to the kink (antikink),
and the specific potential (\ref{V3}) and the solution (\ref{nsol2})
were used in the third line (\ref{dere}).
As expected, the same decent relation (\ref{dece}) is read and the obtained
topological NC kink (antikink) is interpreted as a D$(p-1)$-brane of tension
satisfying ${\cal T}_{p-1}=\pi R{\cal T}_{p}\sqrt{-\alpha_{px}}$
as for each kink
(antikink) in the previous subsection. To confirm let us compute
its NC energy per unit transverse volume
\begin{eqnarray}
\frac{\hat H}{\int dy_{1}\cdots dy_{p-1}}
&=& \int_{-\infty}^{\infty}dx\sqrt{-G}(-{\hat T}^{0}_{~~0})
\nonumber \\
&=& \hat {\cal T}_p \left|\frac{{\hat {\cal T}}_p
{\hat \alpha}_{px}}{\hat T^{11}}\right|
\int_{-\infty}^{\infty} dx
\frac{1}{1 + \left|\frac{{\hat {\cal T}}_p^{2}{\hat \alpha}_{px}}{R^{2}
(T^{11})^2 {\hat \beta}_{p}}\right|x^{2}}
\label{lor}\\
&=& \pi R \hat {\cal T}_p \sqrt{-\hat\alpha_{px}} = {\cal T}_{p-1}.
\label{tenB}
\end{eqnarray}

Though existence of the thick topological NC kink is guaranteed
for any runaway tachyon potential (\ref{vbd}), various special
features involving the exact solution (\ref{nsol2}), the exact
tension formula (\ref{topc})--(\ref{dere}), Lorentzian
distribution of NC energy density (\ref{lor}), tension as NC energy per
unit transverse volume (\ref{tenB}), are collection of evidences
showing BPS nature of the topological NC kink with non-zero
thickness under the specific tachyon potential (\ref{V3}). Even
though $\hat {\cal T}_p$ and ${\hat T}^{11}$ are remained to be
finite, ${\cal T}_p$ (\ref{ret}) and $T^{11}$ (\ref{txr}) are
vanishing due to ${\hat \beta}_{p}=0$. This limit corresponds to
zero thickness limit with keeping BPS nature. Therefore thick NC
kink requires infinite ${\hat T}^{11}$ with keeping
$T^{11}=\sqrt{{\hat \beta}_{p}}\, {\hat T}^{11}$ finite. Since
commutative limit in NCFT corresponds to no DBI type $B_{\mu\nu}$
in EFT where no single topological kink of this subsection exists
except that with zero thickness in
EFT~\cite{Lambert:2003zr,Kim:2003in}, the thick topological NC
kink cannot survive in the commutative limit.

Comparison between the NCFT results and the BCFT results can easily be
made for this single topological NC kink by expanding the sine function
in both $\tilde f$ (\ref{tfx}) and $f$ (\ref{fx}) and keeping the leading
term proportional to $x$, i.e., $\sin^{2}(\sqrt{G_0}\, x/\sqrt{2})
\approx G_0 x^{2}/2$. Then the same discussion for the array of NC kink and
NC antikink is still sustained for the single topological NC kink.

\subsection{Topological NC kink for $\hat U_p(0) < 0$ and $\hat {\cal E}_p >0$}

When $\hat {\cal E}_p > 0$ ($\hat {\cal T}_p^2 <0$ $\&$ $\hat\alpha_{px} >0$),
we have a solution
\begin{equation}\label{nsol3}
\sinh \left(\frac{\hat T(x)}{R}\right)
= \pm \sqrt{1 + \hat u^2}\,
\sinh\left(\frac{x}{{\hat \zeta}}\right).
\end{equation}
The obtained configuration is also a topological kink (antikink)
with a finite asymptotic
slope $\hat T'(\pm\infty) = \pm R/\hat\zeta$.
For this solution
the action (\ref{sia2}) per unit time and transverse volume is expressed by
\begin{eqnarray}\label{sia5}
\frac{\hat S}{-\int dt \int dy_{1}\cdots dy_{p-1}}
&=& \hat {\cal T}_p \sqrt{\hat\beta_{p}(-\hat u^2)}\,
\int_{-\infty}^{\infty} dx\, \hat V^2(\hat T(x))
\nonumber \\
&=&\hat {\cal T}_p \sqrt{\hat\beta_{p}(-\hat u^2)}\,
\int_{-\infty}^{\infty} dx \frac{1}{1 + (1 + \hat u^2) \sinh^2
(x/{\hat \zeta})}
\nonumber \\
&=& 2i\hat {\cal T}_p R
\sqrt{\hat\alpha_{px}} \tan^{-1}\hat u\,.
\end{eqnarray}
Since the quantity $\hat{\cal T}_p$ is imaginary as explained
in Eq.~(\ref{abt}), the resulting action (\ref{sia5}) or equivalently
tension of D$(p-1)$-brane is real.
As explained in Eq.~(\ref{abt}), the quantity $\hat{\cal T}_p$ is
imaginary but the resulting action (\ref{sia5}) is real and is chosen to be
negative, i.e., $\hat {\cal T}_p = -i |\hat {\cal T}_p|$.
So does NC energy density
\begin{equation}\label{ncham1}
-\sqrt{-G}\,\hat T^{0}_{~~0}=
i\hat {\cal T}_p \sqrt{\hat\beta_{p}\hat u^2}\,\hat V^2(\hat T(x))
= \frac{i\hat {\cal T}_p \sqrt{\hat\beta_{p}\hat u^2}}{1
+ (1 + \hat u^2) \sinh^2 (x/{\hat \zeta})}.
\end{equation}
It is localized near the origin and then thin limit ($\hat T^{11} \to 0^-$)
of it can smoothly be taken.

\subsection{NC bounce for $0<\hat {\cal E}_p< \hat U_p(0)$}

When $ 0 < \hat U_p(0)$, $\hat\alpha_{px} > 0$ and then
negative action per unit time and unit
transverse volume is read from Eq.~(\ref{sia3})
\begin{eqnarray}\label{hreb}
\frac{{\hat S}}{-\int dt\int dy_{1}\cdots dy_{p-1}}
&=& \hat{{\cal T}}_{p}\sqrt{\hat\alpha_{px}}\int d\tilde{x}\,
\hat V(\hat T)\sqrt{1-\left(\frac{d\hat T}{d\tilde x}\right)^2}\,,
\end{eqnarray}
where $\tilde x \equiv \sqrt{\hat\beta_p /\hat\alpha_{px}}\, x$.
The form of action (\ref{hreb}) of
static NC kinks coincides exactly with that of the
rolling tachyon with or without DBI type electromagnetic coupling
under exchange of time and spatial
coordinates~\cite{Kim:2003he,Kim:2004zq} and, simultaneously with that
of kinks corresponding to composite of D0F1 in the
EFT~\cite{Kim:2003in,Kim:2003ma}. Thus, there exists a one-to-one
correspondence between a rolling tachyon solution of time
evolution in EFT and a kink solution with spatial distribution for
$0 < \hat U_p(0)$ in NCFT.
With this identification, the pressure
$-{\hat {\cal T}}^{11}$ in our system corresponds to the
Hamiltonian density ${\cal H}$ in the rolling tachyon system.
To be specific, let us perform detailed analysis in subsections 3.4--3.6
distinguished by the value of $\hat {\cal E}_p$.

When $ 0 < \hat {\cal E}_p < \hat U_p(0)$ ($\hat {\cal T}_p^2 >0$,\,
$\&$ $\hat\alpha_{px} > 0$), we have
\begin{equation}\label{nsol4}
\sinh\left(\frac{\hat T(x)}{R}\right)
= \pm \sqrt{\hat u^2 -1} \cosh\left(\frac{x}{\hat{\zeta}}\right)
\end{equation}
which is convex up (convex down) with minimum (maximum) value
$\hat T_{{\rm min}}=R \cosh^{-1}\hat u$ ($-\hat T_{{\rm min}}$)
and has finite asymptotic slope at infinity
$\hat T'(\pm\infty) = \pm R/{\hat \zeta}$.
Therefore, this solution describes an NC bounce.

Tension per unit transverse volume of the NC bounce
is computed from
the action (\ref{hreb}) or the NC energy
\begin{eqnarray}\label{ncham2}
\hat {\cal T}_{p-1}&=&\hat {\cal T}_p \sqrt{\hat\beta_{p}\hat u^2}
\int^{\infty}_{-\infty}dx
\,\hat V^2(\hat T(x)) \nonumber\\
&=& \hat {\cal T}_p \sqrt{\hat\beta_{p}\hat u^2}
\int_{-\infty}^{\infty} dx \frac{1}{1 + ( \hat u^2-1) \cosh^2
(x/{\hat\zeta})}
\nonumber \\
&=& 2R\hat {\cal T}_p\sqrt{\hat\alpha_{px}} \tanh^{-1}
\left(\frac{1}{\hat u}\right)
=2R{\cal T}_p\sqrt{\alpha_{px}}\tanh^{-1}\left(\frac{1}{u}\right)
={\cal T}_{p-1},
\label{oou}
\end{eqnarray}
where we used Eq.~(\ref{ret}), Eqs.~(\ref{etoe})--(\ref{txr}),
and the results of EFT in the last line.

Thin limit is achieved by taking
vanishing pressure, $-{\hat T}^{11} \rightarrow 0$. Then ${\hat u}\rightarrow
\infty$ and the solution (\ref{nsol4}) becomes
singular ${\hat T}(x)\stackrel{-{\hat T}^{11}\rightarrow 0}{
\longrightarrow}\pm\infty\cosh(|{\hat \omega}|x)$ in this limit,
however this thin bounce exists in the tensionless limit
$({\cal T}_{p-1}\propto\tanh^{-1}(1/\infty)\to 0)$ as shown in
Eq.~(\ref{oou}).

\subsection{Tensionless NC half-kink for $0<\hat U_p(0)$ and
$\hat {\cal E}_p= \hat U_p(0)$}

When $\hat {\cal E}_p = \hat U_p(0)$ ($\hat {\cal T}_p^2 >0$
$\&$ $\hat\alpha_{px} >0$), we have a trivial ontop solution
$\hat T(0) = 0$ and nontrivial NC half-kink solution connecting
smoothly unstable symmetric vacuum $\hat T(-\infty)=0$ and one of two stable
broken vacua $\hat T(\infty) = \pm\infty$
\begin{equation}\label{nsol5}
\sinh\left(\frac{\hat T(x)}{R}\right) =
\pm \exp\left(\frac{x-x_{0}}{{\hat \zeta}}\right),
\end{equation}
where $x_{0}$ stands for location of the NC half-kink. Note that
the scale factor ${\hat \zeta}$ (\ref{uzeta}) is fixed for a given
$R$ and the solution (\ref{nsol5}) does not have any free
parameter for the given background so that thin limit of the NC
half-kink cannot be taken.

If we naively compute tension of the NC half-kink from the NC energy density,
it includes vacuum NC energy from the unstable vacuum,
proportional to $\int_{x_{0}}^{\infty} dx\, {\hat V}^2(0)$.
Therefore, it is reasonable to subtract the vacuum NC energy when its tension is
computed. The formula for tension
is obtained by inserting the NC half-kink solution (\ref{nsol5}) as given
in the following:
\begin{eqnarray}
\lefteqn{\frac{{\hat S}}{-\int dt\int dy_{1}\cdots dy_{p-1}}
-\hat {\cal T}_p \sqrt{{\hat\beta_{p}}{\hat u^2}}\int_{x_{0}}^{\infty} dx
\, {\hat V^2 (0)}}\nonumber\\
&=& \hat {\cal T}_p \sqrt{{\hat\beta_{p}}{\hat u^2}}
\left\{ \int_{x_{0}}^{\infty}dx\left[ \frac{1}{1+e^{2
(x-x_{0})/{\hat \zeta}}}-1 \right] + \int_{-\infty}^{x_{0}}dx
\frac{1}{1+e^{2(x-x_{0})/{\hat \zeta}}}
\right\}\nonumber\\
&=&0.
\end{eqnarray}
Amazing enough, the NC half-kink is identified as a candidate of
tensionless half-brane (${\rm D}\frac{1}{2}$-brane) with thickness
due to nonvanishing pressure, $-{\hat T}^{11}$. Thus, by an
arbitrarily small perturbation, the zero mode $x_{0}$ may start to
move to the true vacuum by transferring vacuum energy of the
unstable vacuum to its kinetic energy.

\subsection{Hybrid of two NC half-kinks for $0<\hat U_p(0)$
 $<\hat {\cal E}_p$}

When $ 0<\hat U_p(0) < \hat {\cal E}_p$ ($\hat {\cal T}_p^2 >0$
$\&$ $\hat\alpha_{px} > 0$), we have a monotonically increasing
(decreasing) solution from $\hat T(-\infty) = -\infty~(+\infty)$
to $\hat T(\infty) = +\infty~(-\infty)$
\begin{equation}\label{nsol6}
\sinh\left(\frac{\hat T(x)}{R}\right) = \pm \sqrt{1 -\hat u^2}
\sinh\left(\frac{x}{{\hat \zeta}}\right).
\end{equation}
This configuration can be regarded as a hybrid of two half-kinks joined at
the origin.

Since reality condition of the solution (\ref{nsol6}) does not allow
the limit of ${\hat u} \rightarrow \infty$, thin limit cannot be taken
despite of its localized NC energy density
\begin{eqnarray}\label{4ene}
-\sqrt{-G}{\hat T}^{0}_{~~0}=\hat {\cal T}_p \sqrt{{\hat\beta_{p}}{\hat u^2}}
\frac{1}{1 + (1- \hat u^2) \sinh^2
(x/{\hat \zeta})}.
\end{eqnarray}
Integration of the NC energy density (\ref{4ene}) gives tension of the
hybrid of two D$\frac{1}{2}$-branes
\begin{eqnarray}\label{ncham4}
\hat {\cal T}_p \sqrt{{\hat\beta_{p}}{\hat u^2}}
\int_{-\infty}^{\infty} dx \,\hat V^2(\hat T(x))
=2R\hat {\cal T}_p  \sqrt{\hat\alpha_{px}} \tanh^{-1} {\hat u}
=2R{\cal T}_p\sqrt{\alpha_{px}}\tanh^{-1} {u}.
\end{eqnarray}

We obtained three types of NC kinks for $p\ge2$ through subsections
3.4--3.6, which are closely related with those obtained in EFT of a real
tachyon coupled to U(1) gauge field~\cite{Kim:2003in,Kim:2003ma}. As those
solutions are not supported in EFT without DBI type electromagnetism,
the obtained solutions do not exist in commutative limit
($\lim_{B_{\mu\nu\to 0}}\theta^{\mu\nu}\to 0$).

Various other actions of the NC tachyons have been
proposed~\cite{Dasgupta:2000ft,Harvey:2000jt,Gopakumar:2000rw},
so that comparison between the NC kink solutions obtained in this paper
and those from the other NC tachyon actions deserves to be analyzed
but the previous discussion in Ref.~\cite{Banerjee:2004cw} seems enough
for our cases.

\setcounter{equation}{0}
\section{NC Tachyon Kinks with NC U(1) Gauge Field}
In the previous section we obtained all possible regular static NC kink
configurations representing flat D$(p-1)$-branes from an unstable flat
D$p$-brane in the background of constant NS-NS two-form field.
Naturally such unstable D$p$-brane couples to DBI type U(1) gauge field
so that, in EFT, $B_{\mu\nu}$ is replaced by
$B_{\mu\nu}+F_{\mu\nu}$ which is invariant under gauge transformation
of NS-NS two form field on the brane. Therefore, as far as the field
strength $F_{\mu\nu}$ is constant, nothing is changed for spectrum of
the NC kinks under an identification of $B_{\mu\nu}+F_{\mu\nu}$ as the
gauge invariant two-form field. In fact, even if $B_{\mu\nu}$ is constant,
the field strength $F_{\mu\nu}$ needs not to be constant but has
complicated functional dependence of the coordinates in EFT.
In the context of NCFT, the closed string metric $\eta_{\mu\nu}$ and
NS-NS two-form field $B_{\mu\nu}$ are replaced by the open string
metric $G_{\mu\nu}$ (\ref{omet}) and NC parameter $\theta^{\mu\nu}$
(\ref{ncpa}), and then the field strength $F_{\mu\nu}$ is given by
the NC field strength ${\hat F}_{\mu\nu}$.
In this section let us include the NC gauge field ${\hat A}_{\mu}$ resulting
in
\begin{equation}\label{xde}
{\hat F}_{\mu\nu}(x)=\partial_{\mu}{\hat A}_{\nu}
-\partial_{\nu}{\hat A}_{\mu}-i[{\hat A}_{\mu},{\hat A}_{\nu}]_{\ast},
\end{equation}
where single $x$-dependence is consistent with flatness of
codimension-one D-brane and the assumption on the NC tachyon field
(\ref{asta}). Here the NC commutator in Eq.~(\ref{xde}) is defined
by
\begin{equation}\label{nccm}
[A,B]_{\ast}=A\ast B - B\ast A.
\end{equation}

Let us begin with the following action~\cite{Banerjee:2004cw} similar to
Eq.~(\ref{nct})
\begin{equation}\label{ncfa}
\hat{S}_{{*}}= -\frac{{\hat {\cal T}}_p}{2} \int d^{p+1}x\;
\left[\hat{V}(\hat{T}) \ast \sqrt{ -\hat X_{*} } +\sqrt{ -\hat X_{*}
}\ast \hat{V} (\hat{T}) \right],
\end{equation}
where the determinant ${\hat X}_{\hat F}$ involves ${\hat F}_{\mu\nu}$ as
\begin{eqnarray}\label{nX}
\hat X_{*} \equiv {\det}_\ast \left[ G_{\mu\nu} + \hat F_{\mu\nu}
+ \frac{1}{2} \left(\hat D_\mu \hat T \ast \hat D_\nu \hat T +
\hat D_\nu \hat T \ast \hat D_\mu \hat T \right)\right],
\end{eqnarray}
where NC covariant derivative is
\begin{equation}\label{ncde}
{\hat D}_{\mu}=\partial_{\mu}-i[{\hat A}_{\mu},\; ]_{\ast}.
\end{equation}
When the tachyon effective action (\ref{fa}) was derived~\cite{Kutasov:2003er}
and the equivalence between DBI action and NCFT action was
verified~\cite{Seiberg:1999vs}, the condition of slowly varying
fields was assumed
for both the tachyon field as ${\hat \partial}_{\mu}{\hat D}_\nu {\hat T}$
and the gauge field as ${\hat \partial}_\mu {\hat F_{\nu\rho}}=0$.
Therefore, we can employ a rather simple form of the NC
action~\cite{Banerjee:2004cw}
connected smoothly to the DBI type NC action by
Seiberg-Witten~\cite{Seiberg:1999vs}
\begin{equation}\label{ncfa-1}
\hat{S}_{\hat F}= - {\hat {\cal T}}_p \int d^{p+1}x\;
\left[\hat{V}(\hat{T}) \sqrt{- {\hat X}_{\hat F}} + {\cal O} (\partial \hat F,\,
\partial \hat D \hat T)\right],
\end{equation}
where
\begin{equation}
\hat X_{\hat F} = \det (G_{\mu\nu} + \hat F_{\mu\nu}
+ \hat D_\mu \hat T \hat D_\nu \hat T).
\end{equation}
In addition we will show that the kink configurations from Eq.~(\ref{ncfa-1})
are exactly the same as those from Eq.~(\ref{ncfa}), which is likely to
support a kind of universality.

NC tachyon equation and NC gauge equation from the action (\ref{ncfa-1}) are
\begin{eqnarray}
&& D_\mu \left(
\frac{\hat V (\hat T)}{\sqrt{- \hat X}}\,
\hat C^{\mu\nu}_{{\rm S}} \hat D_\nu\hat T\right)
- \left[ \frac{{\hat T}}{R^{2}}\frac{d\hat V(\hat T)}{d\hat T}
\right]_\ast \sqrt{ - \hat X} =0,
\label{ncte} \\
&& \hat D_\mu\left( \frac{\hat V(\hat T)}{\sqrt{{-\hat X}_{\hat F}}} \,
\hat C^{\mu\nu}_{{\rm A}}\right)
+ i \left[ \hat T\, , \, \frac{\hat V(\hat T)}{
\sqrt{{-\hat X}_{\hat F}}} \,
\hat C^{\mu\nu}_{{\rm S}}\hat D_\nu \hat T \right]_\ast = 0,
\label{ncge}
\end{eqnarray}
where ${\hat C^{\mu\nu}_{{\rm S}}}$ and ${\hat C^{\mu\nu}_{{\rm A}}}$ are
symmetric and antisymmetric parts of cofactors, respectively.
Or some of the equations of motion can be replaced by conservation of
NC energy-momentum~\cite{Banerjee:2003vc}
\begin{eqnarray}\label{nccem}
\hat{D}_{\mu} \hat{T}^{\mu\nu}=0
\end{eqnarray}
for calculational simplicity.

Since we assumed $x$-dependence to both the NC tachyon field (\ref{asta})
and the NC field
strength tensor (\ref{xde}), difficulty of solving the equations of motion
(\ref{ncte})--(\ref{ncge}) with Bianchi identity for the NC gauge
field~\cite{Banerjee:2003vc}
\begin{equation}\label{ncbi}
{\hat D}_{\mu}{\hat F}_{\nu\rho}
+{\hat D}_{\nu}{\hat F}_{\rho\mu}
+{\hat D}_{\rho}{\hat F}_{\mu\nu}=0
\end{equation}
is significantly reduced. The simplest case is $p=1$ where the
Bianchi identity (\ref{ncbi}) becomes trivial, so analysis was
made and results were obtained, which are consistent with those of
EFT ~\cite{Banerjee:2004cw}. From now on let us tackle the case of
D2-brane ($p=2$) on which dynamics of the U(1) gauge field is
expected to be much more complicated due to three components of
electromagnetic field strength tensor and three NC parameters.

For a flat D2-brane, the closed string metric is $g_{\mu\nu}=\eta_{\mu\nu}$
and we set components of the NS-NS two-form field $B_{\mu\nu}$ as
\begin{equation}\label{bmn}
B_{01}=E_{1},\quad B_{02}=E_{2},\quad B_{12}=B, \quad
{\rm with}\quad {\bf E}^{2}=E_{1}^{2}+E_{2}^{2},
\end{equation}
and then the open string metric $G_{\mu\nu}$ and the NC parameter
$\theta^{\mu\nu}$ are determined by Eq.~(\ref{omet}) and Eq.~(\ref{ncpa}),
respectively;
\begin{eqnarray}
G_{00}&=&-({1-{\bf E}^{2}}), \qquad
G_{01}=E_2B, \qquad G_{02}=-{E_1B},\nonumber \\
G_{11}&=&1-E_{1}^{2}+B^{2}, \qquad
G_{12}=-{E_1E_2},\qquad
G_{22}={1-E_2^{2}+B^{2}},
\label{Gmn}
\end{eqnarray}
and
\begin{eqnarray}\label{thmn}
\theta^{01}&=&\frac{E_1}{1-{\bf E}^{2}+B^{2}},  \qquad
\theta^{02}=\frac{E_2}{1-{\bf E}^{2}+B^{2}},  \qquad
\theta^{12}=\frac{-B}{1-{\bf E}^{2}+B^{2}}.
\end{eqnarray}
From the relation (\ref{ret}) we have
\begin{eqnarray}
G_{{\rm s}}=(-G)^{1/4}g_{{\rm s}}=\sqrt{1-{\bf E}^{2}+B^{2}}\,
g_{{\rm s}},\qquad
{\hat {\cal T}}_{2}=(-G)^{-1/4}{\cal T}_{2}=
\frac{{\cal T}_{2}}{\sqrt{1-{\bf E}^{2}+B^{2}}}.
\end{eqnarray}

Since we are interested in obtaining flat D1-branes
or D1F1 composites from the flat unstable
D2-brane, all the NC fields of our consideration depend on
$x$-coordinate but do not depend on $y$-coordinate
as given in Eq.~(\ref{asta}) and Eq.~(\ref{xde}).
Suppose we choose a noncovariant gauge for the NC U(1) gauge field as
\begin{eqnarray}\label{gf}
\theta^{1\mu}{\hat A}_{\mu}=0,
\end{eqnarray}
where the time component ${\hat A}_{0}$ is expressed in terms of
${\hat A}_{2}$.
Inserting this into the NC field strength tensor ${\hat F}_{\mu\nu}$,
we obtain
\begin{eqnarray}
\hat F_{01} \equiv \hat E_1 &=& \partial_0\hat A_1 - \partial_1 \hat
A_0 -i[\hat A_0, \hat A_1]_{\ast} = \partial_0\hat A_1
+\frac{\theta^{12}}{\theta^{10}}
\partial_1 \hat A_2+i\frac{\theta^{12}}{\theta^{10}}[\hat A_2, \hat A_1]_{\ast},
\label{fh1} \\
\hat F_{02} \equiv \hat E_2 &=& \partial_0\hat A_2 - \partial_2 \hat
A_0 -i[\hat A_0, \hat A_2]_{\ast} = \partial_0 \hat A_2
+ \frac{\theta^{12}}{\theta^{10}}
\partial_2 {\hat A}_2 ,
\label{fh2} \\
\hat F_{12} \equiv \hat B &=& \partial_1\hat A_2 - \partial_2 \hat
A_1 -i[\hat A_1, \hat A_2]_{\ast}.
\label{fh3}
\end{eqnarray}
Eliminating the commutator term in Eq.~(\ref{fh1}) and
Eq.~(\ref{fh3}), we have two first-order equations of
${\hat A}_{1}$ and ${\hat A}_{2}$ in $t$ and $y$, i.e., they are
Eq.~(\ref{fh2}) and
\begin{eqnarray}
(\theta^{10}\partial_{0}+\theta^{12}\partial_{2}){\hat A}_{1}&=&
\theta^{10}{\hat E}_{1}(x)-\theta^{12}{\hat B}(x).
\label{f1}
\end{eqnarray}
Their general solutions are
\begin{eqnarray}
{\hat A}_{1}&=& \frac{1}{\theta^{12}}\left[c_{1}\theta^{12}t
+(1-c_{1})\theta^{10}y\right]{\hat E}_{1}(x) \nonumber\\
&& -\frac{1}{\theta^{10}}\left[c_{2}\theta^{12}t
+(1-c_{2})\theta^{10}y\right]{\hat B}(x)+g_{1}(x),
\label{ha1}\\
{\hat A}_{2}&=& \frac{1}{\theta^{12}}\left[c_{3}\theta^{12}t
+(1-c_{3})\theta^{10}y\right]{\hat E}_{2}(x) + g_{2}(x),
\label{ha2}
\end{eqnarray}
where $c_{i}$s $(i=1,2,3)$ are arbitrary constants and $g_{i}(x)$s
$(i=1,2)$ are arbitrary real functions of $x$-coordinate.
Substituting ${\hat A}_{1}$ and ${\hat A}_{2}$ into Eq.~(\ref{fh3}),
we have from the terms proportional to $\theta^{20}$
\begin{eqnarray}
c_1 = c_2 = c_3=c.
\end{eqnarray}
The gauge fixing condition (\ref{gf}) gives an expression of ${\hat A}_{0}$;
\begin{eqnarray}\label{ha0}
{\hat A}_{0}&=&-\frac{\theta^{12}}{\theta^{10}}{\hat A}_{2} \nonumber\\
&=&-\frac{1}{\theta^{10}}\left[c\theta^{12}t
+(1-c)\theta^{10}y\right]{\hat E}_{2}(x)
-\frac{\theta^{12}}{\theta^{10}}g_{2}(x).
\end{eqnarray}
Substituting the NC gauge field (\ref{ha1})--(\ref{ha0}) and its field strength
(\ref{xde}) into the Bianchi identity (\ref{ncbi}), we notice disappearance
of arbitrariness
\begin{equation}\label{bi}
(1-\theta^{10}{\hat E}_1+\theta^{12}{\hat B}){\hat E}_2 '
-{\hat E}_2 (-\theta^{10}{\hat E}_1 ' +\theta^{12}{\hat B}')=0,
\end{equation}
which implies that $c$ and $g_{i}$ are remnants of gauge artifacts.
Since Eq.~(\ref{bi}) results in an algebraic relation among NC parameters and
NC field strength tensor
\begin{equation}\label{bi2}
1-\theta^{10}{\hat E}_1+\theta^{12}{\hat B}= c_{4} {\hat E}_2 ,
\end{equation}
again the NC gauge field (\ref{ha1})--(\ref{ha0}) is inserted into
Eq.~(\ref{fh3}) or Eq.~(\ref{fh1}). Then, with the help of the Bianchi
identity (\ref{bi2}), the arbitrary constants, $c$ and $c_{4}$,
and arbitrary functions
$g_{i}(x)$s should satisfy a constraint equation
\begin{eqnarray}
c(1-c_{4}{\hat E}_{2})-\theta^{10}{\hat E}_{1}+
\theta^{12}c_{4}{\hat E}_{2}g_{2}'+\theta^{10}\theta^{12}{\hat E}_{2}g_{1}'=0.
\end{eqnarray}

Since the NC tachyon field depends only on $x$-coordinate as given in
Eq.~(\ref{asta}) and covariant derivative of the NC tachyon is
expressed in terms of constant (NC) parameters and ${\hat E}_{2}$ as
\begin{eqnarray}
\hat D_\mu \hat T =
\hat E_2 \left( \delta_{\mu 0} \theta^{12}+\delta_{\mu 1}c_{4}
-\delta_{\mu 2}\theta^{10}\right){\hat T}^{'},
\end{eqnarray}
every star product between two NC tachyon fields is replaced by an
ordinary product
\begin{eqnarray}\label{tat}
&&\hat T \ast \hat T = \hat T^2, \quad
\hat D_\mu \hat T \ast \hat D_\nu \hat T =
{\hat D}_\mu {\hat T} {\hat D}_\nu {\hat T}.
\end{eqnarray}

Here we easily notice that a set of constant NC field strength tensor,
$(\hat E_1 , \hat E_2 , {\hat B})$, is a solution of the NC Bianchi identity
(\ref{bi}).
In order to show that the constant NC field strength tensor is
unique solution of the NC equations of motion (\ref{ncte})--(\ref{ncge})
under the ansatz (\ref{asta}) and (\ref{xde}),
let us consider the conservation of NC energy-momentum (\ref{nccem}), giving
\begin{eqnarray}
{\hat D}_{\mu}{\hat T}^{\mu\,0}
=& -\sqrt{-G}{\hat E}_2{\hat B}^{'}{\hat \gamma}+\omega_0{\hat \gamma}^{'}&=0,
\label{emcon0}\\
{\hat D}_{\mu}{\hat T}^{\mu\,1}=& \sqrt{-G}{\hat E}_2
{\hat E}_2^{'}{\hat \gamma}+\omega_1{\hat \gamma}^{'}&=0,
\label{emcon1}\\
{\hat D}_{\mu}{\hat T}^{\mu\,2}=& -\sqrt{-G}
{\hat E}_2{\hat E}_1^{'}{\hat \gamma}
+\omega_2{\hat \gamma}^{'}&=0,
\label{emcon2}
\end{eqnarray}
where we set
\begin{equation}\label{cobe}
{\hat \gamma}=\frac{{\hat{\cal T}}_2{\hat V}}{\sqrt{-{\hat X}_{\ast}}}\, ,
\end{equation}
and $\omega_0$, $\omega_1$, and $\omega_2$ are
\begin{eqnarray}
\omega_0 &=&\sqrt{-G}\Big[-{\hat B}+E_{1}^{2}B+E_2B(1-{\bf E}^{2}+B^{2})\,c_4
+(1+B^{2})B\Big]{\hat E}_{2},
\nonumber\\
\omega_1&=&\sqrt{-G}\Big[{\hat E}_2-E_{1}^{2}E_2+E_2B^{2}-(1-E_{2}^{2})
(1-{\bf E}^{2}+B^{2})\,c_4\Big]{\hat E}_{2},
\nonumber\\
\omega_2&=&\sqrt{-G}\Big[-{\hat E}_1-E_1E_2(1-{\bf E}^{2}+B^{2})\,c_4
-E_1B^{2}-(1-E_1^{2})E_1\Big]{\hat E}_2.
\end{eqnarray}
We multiply some coefficients to
Eqs.~(\ref{emcon0})--(\ref{emcon2}) and add all of them. Then we
obtain
\begin{eqnarray}
0&=&-\theta^{10}{\hat D}_{\mu}{\hat T}^{\mu\,2}
+\theta^{12} {\hat D}_{\mu}{\hat T}^{\mu\,0}
+\frac{1-\theta^{10}{\hat E}_{1}+\theta^{12}{\hat B}}{{\hat E}_{2}}
{\hat D}_{\mu}{\hat T}^{\mu\,1}
\nonumber\\
&=&\left[ (1-\theta^{10}{\hat E}_1+\theta^{12}{\hat B}){\hat E}_2 '
-{\hat E}_2 (-\theta^{10}{\hat E}_1 ' +\theta^{12}{\hat B}')
\right]{\hat \gamma}+\epsilon{\hat \gamma}'
\nonumber\\
&=&\epsilon{\hat \gamma}',
\label{ebe}
\end{eqnarray}
where $\epsilon$ is defined by
\begin{eqnarray}
\epsilon&\equiv& -\theta^{10}\omega_{2} +\theta^{12}\omega_{0}
+\frac{1-\theta^{10}{\hat E}_{1}+\theta^{12}{\hat B}}{{\hat E}_{2}}
\omega_{1}\nonumber\\
&=&\frac{{\hat E}_{2}}{1-{\bf E}^{2}+B^{2}}\Big[1-2E_{1}^{2}+E_1^{4}
-E_{2}^{2}-2E_{1}^{2}B^{2}-B^{4}\nonumber\\
&&\;\;\;\;\;-(1-{\bf E}^{2}+B^{2})\,c_4{\Big\{}2E_{1}^{2}E_2
-(1-E_{2}^{2})(1-{\bf E}^{2}+B^{2})c_4{\Big\}}\Big],
\label{epsi}
\end{eqnarray}
and, in the last line (\ref{ebe}), we used the Bianchi identity (\ref{bi}).
Only trivial solutions are provided by $\epsilon=0$, and then equation of
constant ${\hat \gamma}$ should be examined for nontrivial NC tachyon
solutions.
Once ${\hat \gamma}$ is constant, Eqs.~(\ref{emcon0})--(\ref{emcon2})
force constant NC electromagnetic field as follows
\begin{equation}\label{coem}
\frac{{\hat E}_2{\hat B}^{'}}{\omega_0}=
-\frac{{\hat E}_1 {\hat E}_2^{'}}{\omega_1}
=\frac{{\hat E}_2{\hat E}_1^{'}}{\omega_2}
=\frac{{\hat \gamma}^{'}}{{\hat \gamma}} =0.
\end{equation}
Therefore, the solution of constant ${\hat \gamma}$ and constant NC
electromagnetic field ${\hat F}_{\mu\nu}$
is unique solution of the Bianchi identity (\ref{bi})
and the conservation of NC energy-momentum tensor
(\ref{emcon0})--(\ref{emcon2}).
If we insert this constant solution into the equations of NC tachyon
(\ref{ncte}) and NC gauge field (\ref{ncge}),
it automatically satisfies the equations.

The only nontrivial equation to be solved now is that of constant
${\hat \gamma}$ (\ref{cobe}). In order to use the results of the
previous section, let us compute the determinant in
Eq.~(\ref{cobe}). Though every component of ${\hat X}_{\mu\nu}$
contains the term of ${\hat T}^{'2}$, all the quartic terms ${\hat
T}^{'4}$ and sixth-order terms ${\hat T}^{'6}$ vanish but constant
and quadratic terms survive
\begin{eqnarray}
-{\hat X}_{\hat F}&=& -{\rm det}\Big[G_{\mu\nu}+{\hat F}_{\mu\nu}+{\hat E}_2^{2}
(\delta_{\mu0}\theta^{12}+\delta_{\mu1}\;c_4-\delta_{\mu2}\theta^{10})
(\delta_{\nu0}\theta^{12}+\delta_{\nu1}\;c_4-\delta_{\nu2}\theta^{10})
{\hat T}^{'2}\Big]\nonumber\\
&=&-{\rm det}(G_{\mu\nu}+{\hat F}_{\mu\nu})
- \left. {\hat C}_{\hat F}^{\mu\nu}\right|_{{\hat T}^{'}=0}
\left. {\hat X}^{\hat F}_{\mu\nu}\right|_{G_{\mu\nu}={\hat F}_{\mu\nu}=0}
\label{cald}\\
&=&{\hat \beta}_{\hat F}-{\hat\alpha}_{\hat F}
{\hat T}^{'2},
\end{eqnarray}
where ${\hat\beta}_{\hat F}= {\rm det}(G_{\mu\nu}+{\hat F}_{\mu\nu})$ and
explicit form of ${\hat\alpha}_{\hat F}$ is
\begin{eqnarray}
{\hat\alpha}_{\hat F}&=&\left[
(G_{11}G_{22}-G_{12}^{2}+{\hat F}_{12}^{2})(\theta^{12})^{2}
+(G_{00}G_{22}-G_{02}^{2}+{\hat F}_{02}^{2})c_{4}^{2}\right. \nonumber\\
&&\hspace{-0mm}+(G_{00}G_{11}-G_{01}^{2}+{\hat F}_{01}^{2})(\theta^{01})^{2}
-2(G_{01}G_{22}+G_{02}G_{12}-{\hat F}_{02}{\hat F}_{12})c_{4}\theta^{12}
\\
&&\hspace{-0mm}\left.
-2(G_{01}G_{12}+G_{11}G_{02}+{\hat F}_{01}{\hat F}_{12})\theta^{01}\theta^{12}
-2(G_{00}G_{12}-G_{01}G_{02}+{\hat F}_{01}{\hat F}_{02})\theta^{01}c_{4}
\right]{\hat E}_2^{2}.
\nonumber
\end{eqnarray}
For ${\hat X}^{\hat F}_{\mu\nu}|_{G_{\mu\nu}={\hat F}_{\mu\nu}=0}$ in
Eq.~(\ref{cald}), we pretended ${\hat F}_{\mu\nu}=0$ for convenience but
actually ${\hat E}_2(\equiv {\hat F}_{02}) \ne 0$.
Then the equation of constant ${\hat \gamma}$ (\ref{cobe})
is rewritten in a form of
conservation of mechanical energy ${\cal E}_{\hat F}$ of a unit-mass particle
in 1-dimensional motion, of which coordinate is ${\hat T}$, time $x$,
and potential $U_{\hat F}$
\begin{equation}\label{eu3}
{\hat {\cal E}}_{\hat F}({\hat T})
= \frac{1}{2}{\hat T}^{'2}+{\hat U}_{\hat F}({\hat T}),
\end{equation}
where
\begin{equation}\label{hho3}
{\hat {\cal E}}_{\hat F}({\hat T})
=\frac{{\hat\beta}_{\hat F}}{2{\hat\alpha}_{\hat F}},
\qquad {\hat U}_{\hat F}({\hat T})=
\frac{1}{2{\hat\alpha}_{\hat F}
{\hat \gamma}^2}[{\hat{\cal T}}_{2}{\hat V}({\hat T})]^{2}.
\end{equation}

Eq.~(\ref{eu3}) is formally the same first-order equation as Eq.~(\ref{hhoe}).
At the beginning we had six free parameters, namely, three NC parameters
$(\theta^{01},\theta^{02},\theta^{12})$
(or equivalently constant background NS-NS 2-form field
$(E_{1},E_{2},B)$) and three constant NC electromagnetic field
$({\hat E}_{1},{\hat E}_{2},{\hat B})$, and an integration constant
${\hat \gamma}$, however general NC kink solutions of Eq.~(\ref{eu3}) are
classified by two parameters ${\hat {\cal E}}_{\hat F}$ and
${\hat U}_{\hat F}({\hat T}=0)$. Therefore, the analysis in the previous section
is applicable exactly in the same manner. As far as we have runaway
NC tachyon potential (\ref{vbd}), we obtain six types of
NC kink solutions and exact solutions for the $1/\cosh$-potential (\ref{V3})
as shown in Table~1 and Eq.~(\ref{rss3}). Here we do not repeat
the same detailed analysis.

Since antisymmetric part of the cofactor ${\hat C}_{{\hat
F}}^{\mu\nu}$ of ${\hat X}^{{\hat F}}_{\mu\nu}$ does not vanish,
conjugate momentum ${\hat \Pi}^{i}={\hat \gamma}{\hat C}_{{\hat
F}}^{0i}$ of the NC gauge field ${\hat F}_{\mu\nu}$ also does not
vanish
\begin{eqnarray}
{\hat \Pi}^{1}={\hat \gamma}{\hat C}_{{\hat F}{\rm A}}^{01}&=&
{\hat \gamma}( G_{20}{\hat B}-G_{21}{\hat E}_{2}+G_{22}{\hat E}_{1}
-\theta^{01}{\hat E}_{2}^{2}{\hat T}^{'2}),
\label{pi1}\\
{\hat \Pi}^{2}={\hat \gamma}{\hat C}_{{\hat F}{\rm A}}^{02}&=&
{\hat \gamma}( -G_{10}{\hat B}-G_{11}{\hat E}_{2}-G_{12}{\hat E}_{1}
+c_{4}{\hat E}_{2}^{2}{\hat T}^{'2}),
\label{pi2}
\end{eqnarray}
where constant ${\hat \gamma}$ is given in Eq.~(\ref{cobe}). Both
${\hat \Pi}^{1}$ and ${\hat \Pi}^{2}$ involve localized piece
proportional to ${\hat T}^{'2}$ near the position of D1-brane
$x=0$ in addition to constant part. The constant part can be
interpreted as constant F1 fluid
density~\cite{Gibbons:2000hf,Gibbons:2002tv,Kwon:2003qn} and the localized piece
stands for confined F1 along D1~\cite{Kim:2003in,Kim:2003ma}.
Therefore, the obtained objects are identified with D1F1 bound in
the background of constantly distributed F1s, i.e., they are array
of D1F1-${\bar {\rm D}}$1F1, single D1F1, half-DF1, and their
composites.

When the DBI type NC electromagnetic field is added, the resultant
NC action (\ref{ncfa}) is proven to be equivalent to that of commutative
EFT only up to the leading NC parameter
${\cal O}(\theta^{\mu\nu})$~\cite{Banerjee:2004cw}.
A noteworthy observation of this section is the fact that spectrum of
the NC kinks qualitatively coincides with that of commutative EFT.
So it might be intriguing to study the relation between the NCFT action
and the commutative EFT action further.

\section{Summary and Discussion}

In this paper we considered kink (domain wall) solutions in various NCFTs.
For the NCFT of a real scalar field in flat metric, we showed that all the
kink solutions in ordinary scalar field theory with quadratic derivative term
and polynomial scalar potential become NC kink solutions in NCFT.
In $\phi^{4}$-theory the obtained decent relation, given by
the ratio between the local maximum of the potential and energy of
the NC kink, has about 20$\%$ difference
in its coefficient from that of string theory.

We studied DBI type effective action of tachyon field with
constant open string metric and NC parameter, and obtained all
possible static NC kink solutions classified by array of NC
kink-antikink, BPS or non-BPS single topological NC kink,
tensionless NC half-kink, NC bounce, and hybrid of two half-kinks.
For specific $1/\cosh$ type NC tachyon potential, those are given
by exact solutions and subsequently decent relation between
tension of an unstable D$p$-brane and stable D$(p-1)$-brane is
also reproduced in the same as that of EFT and BCFT, which
supports identification of the unit NC kink as a stable
codimension-one D-brane. When DBI type U(1) NC gauge field is
turned on on an unstable D2-brane, gauge equation and NC Bianchi
identity dictate that every field strength component should be
constant. Therefore, forms of the obtained solutions are the same
as those without the NC gauge field, but they carry F1 charge
localized on the codimension-one D-brane. Extension to the case of
arbitrary $p$ is extremely complicated but is likely to result in
the same constant field strength of NC gauge field.

UV-IR mixing is a representative property in
NCFT~\cite{Minwalla:1999px} and it is realized in GMS
solitons~\cite{Gopakumar:2000zd}, however it needs further study
to answer the question whether or not NC tachyon kinks share such
property. We obtained all possible kinks interpreted as flat
codimension-one branes, and NC tachyon tubes from tachyon
tubes~\cite{Kim:2003uc} are worth being reproduced as for thin
objects~\cite{Hashimoto:2000mt}. Since NCFT is an effective
theory, it is definitely intriguing that we obtain those
configurations in the context of (boundary) string field
theory~\cite{Witten:2000nz}.

\section*{Acknowledgments}
The authors would like to thank D. Bak, R. Banerjee, C. Kim
for helpful discussions and
A. Sen for valuable suggestion.
This work is the result of research activities
(Astrophysical Research Center for
the Structure and Evolution of the Cosmos (ARCSEC)) supported by
Korea Science $\&$ Engineering Foundation.

\end{document}